\documentclass[11pt,a4paper]{article}
\pdfoutput=1
\usepackage[utf8]{inputenc}
\usepackage{jheppub}
\hypersetup{unicode=true,bookmarksopen=true}

\usepackage{amsthm}
\usepackage{mathtools}
\usepackage{amsthm}
\usepackage{lmodern}
\usepackage[T1]{fontenc}
\usepackage{textalpha}
\usepackage{bbm}
\DeclareMathAlphabet{\mathbfi}{OML}{cmm}{b}{it}

\let\originalleft\left
\let\originalright\right
\renewcommand{\left}{\mathopen{}\mathclose\bgroup\originalleft}
\renewcommand{\right}{\aftergroup\egroup\originalright}
\makeatletter
\newcommand{\biggg}{\bBigg@\thr@@}
\newcommand{\Biggg}{\bBigg@{3.5}}
\makeatother

\makeatletter
\newenvironment{equations}[1][]{\subequations\ifx\relax#1\relax\else\label{#1}\fi\align\ignorespaces}{\endalign\ignorespacesafterend\endsubequations}
\def\@spliteq#1{\begin{equation}\begin{split}#1\end{split}\end{equation}}
\def\@spliteqstar#1{\begin{equation*}\begin{split}#1\end{split}\end{equation*}}
\def\splitequation{\collect@body\@spliteq}
\expandafter\def\csname splitequation*\endcsname{\collect@body\@spliteqstar}

\expandafter\def\csname endsplitequation*\endcsname{\ignorespacesafterend}
\makeatother

\makeatletter
\g@addto@macro\@floatboxreset\centering
\makeatother

\newcommand{\eqend}[1]{\,#1}
\renewcommand{\vec}[1]{{\ifnum9<1#1\mathbf{#1}\else\ifcat\noexpand#1\relax\boldsymbol{#1}\else\mathbfi{#1}\fi\fi}}
\newcommand{\mathe}{\mathrm{e}}
\newcommand{\mathi}{\mathrm{i}}
\newcommand{\total}{\mathop{}\!\mathrm{d}}

\newcommand{\abs}[1]{{\left\lvert{#1}\right\rvert}}

\newcommand{\sgn}{\operatorname{sgn}}

\newcommand{\tr}{\operatorname{tr}}
\newcommand{\bigo}[1]{\mathcal{O}\left({#1}\right)}
\makeatletter
\def\bra{\@ifnextchar[\bra@size\bra@nosize}
\def\bra@size[#1]#2{#1\langle{#2}#1\rvert}
\def\bra@nosize#1{\left\langle{#1}\right\rvert}
\def\ket{\@ifnextchar[\ket@size\ket@nosize}
\def\ket@size[#1]#2{#1\lvert{#2}#1\rangle}
\def\ket@nosize#1{\left\lvert{#1}\right\rangle}
\makeatother

\newcommand{\normord}[1]{\mathopen{:}{#1}\mathclose{:}}
\newcommand{\supp}{\operatorname{supp}}
\newcommand{\artanh}{\operatorname{artanh}}

\newcommand{\phiclass}{\overline{\raisebox{0.5em}{}\varphi}}

\makeatletter
\gdef\@fpheader{\strut}
\makeatother

\allowdisplaybreaks

\begin{document}


\title{Relative entropy for \texorpdfstring{$\lambda \phi^4$}{\textlambda\textphi\textfoursuperior} in the Rindler wedge}

\author[1]{Markus B. Fröb}
\author[2,3]{\!\!,\ Albert Much}
\author[4]{and Kyriakos Papadopoulos}

\affiliation[1]{Department Mathematik, Friedrich-Alexander-Universit{\"a}t Erlangen-N{\"u}rnberg, Cauerstra{\ss}e 11, 91058 Erlangen, Germany}

\affiliation[2]{Institut f{\"u}r Theoretische Physik, Universit{\"a}t Leipzig, Br{\"u}derstra{\ss}e 16, 04103 Leipzig, Germany}

\affiliation[3]{Department of Applied Physics and Astronomy, College of Sciences, University of Sharjah, Sharjah, United Arab Emirates}

\affiliation[4]{Department of Mathematics, College of Science, Kuwait University, Sabah Al Salem University City, P.O.~Box 5969, Safat 13060, Shadadiya, Kuwait}

\emailAdd{markus.froeb@fau.de}
\emailAdd{amuch@sharjah.ac.ae}
\emailAdd{kyriakos.papadopoulos@ku.edu.kw}

\abstract{We consider the relative entropy between the vacuum and a coherent state in the Rindler wedge for an interacting $\lambda \phi^4$ theory to first order in $\lambda$. We construct the perturbatively interacting Weyl algebra of the wedge, and employ Tomita--Takesaki modular theory and the Araki--Uhlmann formula to compute the relative entropy. We verify that the relative entropy reduces to the classical (interacting) boost Noether charge, analogously to the free theory, and that the Bekenstein bound holds.}

\keywords{}

\maketitle

\section{Introduction}

Relative entropy $S_\mathrm{rel}$ is a powerful tool to quantify the difference between states in quantum theory, with a precise operational definition coming from quantum information theory~\cite{vedral2002}. In quantum mechanics, it is defined as $S_\mathrm{rel}(\rho \vert \sigma) = \tr\left( \rho \ln \rho - \rho \ln \sigma \right)$ for two density matrices $\rho$ and $\sigma$, while the von Neumann entropy of a single state is given by $S_\mathrm{vN}(\rho) = - \tr\left( \rho \ln \rho \right)$. If the density matrices $\rho$ and $\sigma$ are obtained by starting with a global pure state and tracing out the degrees of freedom outside of a given region $R$, giving the reduced density matrices $\rho_R$ and $\sigma_R$, the von Neumann entropy becomes the entanglement entropy $S_\mathrm{EE}(R) = S_\mathrm{vN}(\rho_R)$ and the relative entropy can be thought of as relative entanglement entropy. However, in contrast to $S_\mathrm{EE}$ which is UV-divergent in quantum field theory, relative entropy is a well-defined and UV-finite quantity~\cite{marolfwall2016}. This can be seen by employing Tomita--Takesaki theory~\cite{tomita1967,takesaki1970}, where it can be computed using the Araki--Uhlmann formula~\cite{araki1975,araki1976,uhlmann1977}
\begin{equation}
S_\mathrm{rel}(\omega \vert \phi) = - \omega\left( \ln \Delta_{\phi, \omega} \right) \eqend{.}
\end{equation}
In this formula, $\omega$ and $\phi$ are two faithful normal states on a von Neumann algebra $\mathcal{A}$, and $\ln \Delta_{\phi, \omega}$ is the relative modular Hamiltonian associated to these states and the algebra $\mathcal{A}$. In quantum mechanics, writing $\omega(a) = \tr\left( \rho a \right)$ and $\phi(a) = \tr\left( \sigma a \right)$ for $a \in \mathcal{A}$ it is easy to see that the Araki--Uhlmann formula recovers the quantum-mechanical expression, using that in this case the relative modular operator is given by $\Delta_{\phi, \omega} = L(\sigma) R(\rho^{-1})$, with $L/R$ denoting multiplication from the left/right:
\begin{splitequation}
S_\mathrm{rel}(\omega \vert \phi) &= - \omega\left( \ln \Delta_{\phi, \omega} \right) = - \tr\left( \rho^\frac{1}{2} \ln \Delta_{\phi, \omega} \, \rho^\frac{1}{2} \right) = - \tr\left( \rho^\frac{1}{2} \Bigl[ L(\ln \sigma) - R(\ln \rho) \Bigr] \rho^\frac{1}{2} \right) \\
&= - \tr\left( \rho \ln \sigma - \rho \ln \rho \right) = S_\mathrm{rel}(\rho \vert \sigma) \eqend{.}
\end{splitequation}

However, in general it is very difficult to determine the relative modular Hamiltonian. One exception is when $\phi$ is a unitary excitation of a (cyclic and separating) vector state $\omega = \bra{\Omega} \cdot \ket{\Omega}$, such that $\phi(a) = \bra{\Omega} U^\dagger a U \ket{\Omega}$ with a unitary $U \in \mathcal{A}$. In this case, it is known~\cite{araki1974,arakimasuda1982,casinigrillopontello2019,longo2019,lashkari2019,lashkariliurajagopal2021} that
\begin{equation}
\label{eq:delta_rel_unitary}
\Delta_{\phi, \omega} = U \Delta_{\omega} \, U^\dagger \eqend{,}
\end{equation}
and hence
\begin{equation}
\label{eq:s_rel_unitary}
S_\mathrm{rel}(\omega \vert \phi) = - \bra{\Omega} U \ln \Delta_{\omega} \, U^\dagger \ket{\Omega} \eqend{,}
\end{equation}
where $\Delta_{\omega}$ is the modular operator of the state $\omega$, and its logarithm $\ln \Delta_{\omega}$ is the modular Hamiltonian. In turn, this operator is known in various situations\footnote{We refer to Ref.~\cite{froeb2023} for a list of analytically known modular Hamiltonians, as well as the more recent~\cite{cadamurofroebminz2025,cadamurofroebpereznadal2024,caminiticapecciaciambellimyers2025,tonnitrezzi2026} and further references therein.}, with the most famous one given by the Bisognano--Wichmann theorem~\cite{bisognanowichmann1975,bisognanowichmann1976}. Namely, for $\mathcal{A}$ the algebra of fields in the Rindler wedge $\mathcal{W}_\mathrm{R} = \{ x \in \mathbb{R}^{1,d-1} \colon x^1 \geq \abs{x^0} \}$ and $\omega$ the Minkowski vacuum state, the modular Hamiltonian is proportional to the generator of boosts: $\ln \Delta_{\omega} = 2 \pi \mathcal{M}^{01}$. While this result holds for any Wightman QFT, even interacting, many other modular Hamiltonians are only known for free theories or for CFTs where the enhanced symmetry simplifies computations, and hence also relative entropy has been mostly studied in these cases.

For (perturbatively) interacting theories, comparably few explicit results are known. For example, for a CFT perturbed by a relevant operator one can relate the relative entropy between vacuum states of the free and interacting theories to the change in entanglement entropy which can be explicitly computed~\cite{casinitestetorroba2017,rosenhaussmolkin2015}. Explicit perturbative expressions are also known for the relative entropy between thermal states of the free and interacting scalar field~\cite{dragofaldinopinamonti2018b}, or for fermions perturbed by an external electromagnetic field between equilibrium states of the unperturbed and the perturbed theory~\cite{brunettifredenhagenpinamonti2025}. Moreover, a perturbative series for the relative modular Hamiltonian is known if both states can be written as perturbative excitations of a common state~\cite{lashkariliurajagopal2021,lashkariliurajagopal2023}.

In this work, we consider the relative entropy between the Minkowski vacuum and a coherent excitation thereof, but for a perturbatively interacting theory that is not conformal. To our knowledge, this situation has not been considered yet, and we study $\lambda \phi^4$ theory in the Rindler wedge as the simplest example. As a first step, we review the classical theory in section~\ref{sec:classical}, determine the solution $\varphi^\mathrm{int}_f$~\eqref{eq:phiclass_int_def} of the interacting Klein--Gordon equation which in the free theory is given by $\varphi_f^{(0)}(x) = \int \Delta(x,y) f(y) \total^d y$ with the commutator function $\Delta$ and an arbitrary compactly supported function $f$, and which evolves causally in the sense that $\supp \varphi^\mathrm{int}_f \subseteq J^+(\supp f) \cup J^-(\supp f)$, and finally compute the classical Noether charges. In section~\ref{sec:quantum}, we quantize the theory up to linear order in $\lambda$ and compute the renormalized interacting field operator $\Phi(f)$~\eqref{eq:phi_int_f}, stress tensor $\mathcal{T}_{\mu\nu}$~\eqref{eq:stress_free}, \eqref{eq:stress_order1} and Noether charges~\eqref{eq:minkowski_lorentz_generators}. We then show how to construct the interacting von Neumann algebra $\mathcal{A}$ of the Rindler wedge $\mathcal{W}_\mathrm{R}$, which is composed of interacting Weyl operators $W(f) = \mathe^{\mathi \Phi(f)}$~\eqref{sec:quantum_weyl} depending on a function $f$ with compact support inside the wedge, $\supp f \subset \mathcal{W}_\mathrm{R}$. In particular, we show that the interacting Weyl operators $W(f)$ and $W(g)$ commute if $f$ and $g$ are spacelike separated, even though the interacting field $\Phi(f)$ involves integrations over the causal past $J^-(\supp f)$ extending all the way to past infinity~\eqref{eq:phi_int_f}, in accordance with the general results in algebraic QFT~\cite{duetschfredenhagen2001}.

We proceed in section~\ref{sec:entropy} with the computation of the relative entropy $S_\mathrm{rel}(\Omega \Vert W(f) \Omega)$ between the (interacting) Minkowski vacuum $\ket{\Omega}$ and the interacting coherent state \linebreak $W(f) \ket{\Omega}$, with $f$ supported in the Rindler wedge. Using the relation~\eqref{eq:s_rel_unitary} and the Bisognano--Wichmann theorem, the computation boils down to the computation of the expectation value of the boost generator $\mathcal{M}^{01}$, obtained from the interacting stress tensor $\mathcal{T}_{\mu\nu}$, in the state $W(-f) \ket{\Omega}$. Analogous to the free theory~\cite{casinigrillopontello2019,froebmuchpapadopoulos2023}, the result is given by the classical Noether charge associated to boosts, evaluated on a solution of the interacting Klein--Gordon equation, and we show that the required solution is exactly the causal one $\varphi^\mathrm{int}_f$ that we computed previously. Finally, we prove that our result fulfills a rigorous version of the Bekenstein bound~\eqref{eq:srel_bekenstein}, and conclude in section~\ref{sec:conclusion}.

\section{Classical \texorpdfstring{$\lambda \varphi^4$}{\textlambda\textphi\textfoursuperior} theory}
\label{sec:classical}

We consider the classical (minimally coupled) action of a free scalar field with mass $m$ and coupling constant $\lambda > 0$ in $d$ dimensions
\begin{equation}
S = - \frac{1}{2} \int \left( \nabla_\mu \varphi \nabla^\mu \varphi + m^2 \varphi^2 + \frac{\lambda}{12} \varphi^4 \right) \sqrt{-g} \total^d x \eqend{,}
\end{equation}
whose variations yield the classical (minimally coupled) stress tensor
\begin{equation}
\label{eq:stress_tensor}
T_{\mu\nu}(\varphi) = \nabla_\mu \varphi \nabla_\nu \varphi - \frac{1}{2} g_{\mu\nu} \nabla^\rho \varphi \nabla_\rho \varphi - \frac{1}{2} m^2 g_{\mu\nu} \varphi^2 - \frac{\lambda}{4!} g_{\mu\nu} \varphi^4
\end{equation}
and the classical equation of motion
\begin{equation}
\label{eq:eom_class}
E(\varphi) = \frac{\delta S}{\delta \varphi} = \left( \nabla^2 - m^2 \right) \varphi - \frac{\lambda}{6} \varphi^3 \eqend{.}
\end{equation}
As is well known, the divergence of the stress tensor $\nabla^\mu T_{\mu\nu}(\varphi) = E(\varphi) \nabla_\nu \varphi$ vanishes when the equation of the motion is fulfilled. The assumption $\lambda > 0$ ensures that the classical energy $T_{00}$ is positive and the theory is stable.

\begin{figure}[t]
\label{fig:causalsupport}
\includegraphics[width=0.8\textwidth]{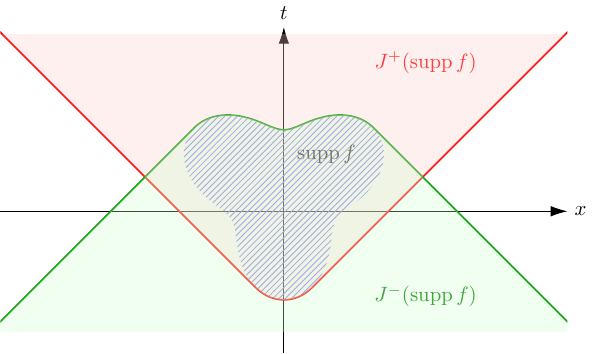}
\caption{The causal future and past $J^\pm(\supp f)$ of a compactly supported function $f$.}
\end{figure}

Specializing to flat space, in the free theory there are unique retarded and advanced Green's functions satisfying
\begin{equation}
\label{eq:gretadv_fundamentalsol}
\left( \partial^2 - m^2 \right) G^\mathrm{ret}(x,y) = \left( \partial^2 - m^2 \right) G^\mathrm{adv}(x,y) = \delta^d(x-y)
\end{equation}
and, with the definitions
\begin{equation}
\label{eq:gret_f_def}
( G^\mathrm{ret} f )(x) = \int G^\mathrm{ret}(x,y) f(y) \total^d y
\end{equation}
and analogously for $G^\mathrm{adv} f$, having the support conditions
\begin{equation}
\label{eq:gretadv_support}
\supp( G^\mathrm{ret} f ) \subseteq J^+(\supp f) \eqend{,} \quad \supp( G^\mathrm{adv} f ) \subseteq J^-(\supp f)
\end{equation}
with $J^\pm$ the causal future or past~\cite{baerginouxpfaeffle2007}, see Fig.~\ref{fig:causalsupport}.
Explicitly, in Fourier space they read
\begin{splitequation}
\label{eq:gretadv_def}
G^\mathrm{ret}(x,y) &= G^\mathrm{adv}(y,x) = - \lim_{\epsilon \to 0^+} \int \frac{\mathe^{\mathi p (x-y)}}{- \left( p^0 + \mathi \epsilon \right)^2 + \omega_\vec{p}^2} \frac{\total^d p}{(2\pi)^d} \\
&= - \Theta(x^0-y^0) \int \frac{\sin\left[ \omega_\vec{p} (x^0-y^0) \right]}{\omega_\vec{p}} \mathe^{\mathi \vec{p} (\vec{x}-\vec{y})} \frac{\total^{d-1} p}{(2\pi)^{d-1}}
\end{splitequation}
with $\omega_\vec{p} = \sqrt{ \vec{p}^2 + m^2 }$. We also define the commutator function (also called causal propagator)
\begin{splitequation}
\label{eq:delta_def}
\Delta(x,y) = G^\mathrm{ret}(x,y) - G^\mathrm{adv}(x,y) &= - 2 \pi \mathi \int \delta(p^2+m^2) \sgn(p^0) \mathe^{\mathi p (x-y)} \frac{\total^d p}{(2\pi)^d} \\
&= - \int \frac{\sin\left[ \omega_\vec{p} (x^0-y^0) \right]}{\omega_\vec{p}} \mathe^{\mathi \vec{p} (\vec{x}-\vec{y})} \frac{\total^{d-1} p}{(2\pi)^{d-1}} \eqend{,}
\end{splitequation}
which is a bisolution of the Klein--Gordon equation, $\partial_x^2 \Delta(x,y) = \partial_y^2 \Delta(x,y) = 0$, and the time-symmetric Green's function (also called Dirac propagator)
\begin{equation}
\label{eq:overlinedelta_def}
\overline{\Delta}(x,y) = \frac{1}{2} G^\mathrm{ret}(x,y) + \frac{1}{2} G^\mathrm{adv}(x,y) \eqend{.}
\end{equation}
$\Delta f$ and $\overline{\Delta} f$ are defined in the same way as for the retarded propagator~\eqref{eq:gret_f_def}.

The interacting equation of motion $E(\varphi) = 0$~\eqref{eq:eom_class} can then be solved perturbatively: given a compactly supported function $f \in C_0^\infty(\mathbb{R}^{1,d-1})$, the field
\begin{equation}
\varphi_f^{(0)}(x) = ( \Delta f )(x)
\end{equation}
solves the Klein--Gordon equation since $\Delta$ is a bisolution. The corresponding retarded and advanced solutions $\varphi_f^\mathrm{ret/adv}$ fulfill
\begin{equation}
\varphi_f^\mathrm{ret/adv}(x) = \varphi_f^{(0)}(x) + \frac{\lambda}{6} \int G^\mathrm{ret/adv}(x,y) \left[ \varphi_f^\mathrm{ret/adv}(y) \right]^3 \total^d y \eqend{,}
\end{equation}
but we also have the time-symmetric solution that fulfills
\begin{equation}
\phiclass_f(x) = \varphi_f^{(0)}(x) + \frac{\lambda}{6} \int \overline{\Delta}(x,y) \left[ \phiclass_f(y) \right]^3 \total^d y \eqend{.}
\end{equation}
In all cases, series expansions in $\lambda$ are obtained by repeatedly substituting the left-hand side into the right-hand side:
\begin{splitequation}
\varphi_f^\mathrm{ret}(x) &= \varphi_f^{(0)}(x) + \frac{\lambda}{6} \int G^\mathrm{ret}(x,y) \left[ \varphi_f^{(0)}(y) + \frac{\lambda}{6} \int G^\mathrm{ret}(y,z) \varphi_f^\mathrm{ret}(z) \total^d z \right]^3 \total^d y \\
&= \varphi_f^{(0)}(x) + \frac{\lambda}{6} \int G^\mathrm{ret}(x,y) \left[ \varphi_f^{(0)}(y) \right]^3 \total^d y \\
&\qquad+ \frac{\lambda^2}{12} \iint G^\mathrm{ret}(x,y) G^\mathrm{ret}(y,z) \left[ \varphi_f^{(0)}(y) \right]^2 \varphi_f^{(0)}(z) \total^d z \total^d y + \bigo{\lambda^3} \eqend{,}
\end{splitequation}
and analogously for $\varphi_f^\mathrm{adv}$ and $\phiclass_f(x)$.

Of course also $f$ may depend on $\lambda$ in some complicated way, which is in fact useful to ensure desired support properties of the solution. Namely, from the definition of $\Delta$ we obtain
\begin{equation}
\supp( \Delta f ) \subseteq \supp( G^\mathrm{ret} f ) \cup \supp( G^\mathrm{adv} f ) = J^+(\supp f) \cup J^-(\supp f) \eqend{,}
\end{equation}
and thus already for the first perturbative correction
\begin{equation}
\varphi_f^\mathrm{ret,(1)}(x) = \left. \frac{\partial}{\partial \lambda} \varphi_f^\mathrm{ret}(x) \right\rvert_{\lambda = 0} = \frac{1}{6} \int G^\mathrm{ret}(x,y) \bigl[ ( \Delta f )(y) \bigr]^3 \total^d y
\end{equation}
we cannot say anything more than $\supp \varphi_f^\mathrm{ret,(1)} \subseteq \mathbb{R}^{1,d-1}$; the same applies to $\varphi_f^\mathrm{adv,(1)}$ and $\phiclass_f^{(1)}$. However, by suitably correcting $f$, we can do better. Consider $g \in C_0^\infty(\mathbb{R}^{1,d-1})$ with $\supp g \cap [ J^+(\supp f) \cup J^-(\supp f) ] = \emptyset$, i.e., the supports of $f$ and $g$ are spacelike separated. Then we have
\begin{equation}
\int g(x) \varphi_f^{(0)}(x) \total^d x = \int g(x) ( \Delta f )(x) \total^d x = 0
\end{equation}
because $\Delta f$ vanishes on the support of $g$. We can achieve the same at least for the first perturbative correction, and compute
\begin{splitequation}
&\int g(x) \varphi_f^\mathrm{ret,(1)}(x) \total^d x = \frac{1}{6} \iint g(x) G^\mathrm{ret}(x,y) \bigl[ ( \Delta f )(y) \bigr]^3 \total^d y \total^d x \\
&\quad= \frac{1}{6} \iiint g(x) G^\mathrm{ret}(x,y) \left[ G^\mathrm{ret}(y,z) - G^\mathrm{adv}(y,z) \right] f(z) \bigl[ ( \Delta f )(y) \bigr]^2 \total^d z \total^d y \total^d x \\
&\quad= \frac{1}{6} \iiint g(x) G^\mathrm{ret}(x,y) G^\mathrm{ret}(y,z) f(z) \bigl[ ( \Delta f )(y) \bigr]^2 \total^d z \total^d y \total^d x \\
&\qquad- \frac{1}{6} \iiint g(x) G^\mathrm{adv}(x,y) G^\mathrm{adv}(y,z) f(z) \bigl[ ( \Delta f )(y) \bigr]^2 \total^d z \total^d y \total^d x \\
&\qquad- \frac{1}{6} \iiint g(x) \Delta(x,y) G^\mathrm{adv}(y,z) f(z) \bigl[ ( \Delta f )(y) \bigr]^2 \total^d z \total^d y \total^d x \eqend{,}
\end{splitequation}
where we employed twice the definition~\eqref{eq:delta_def} of $\Delta$. For the first integral to be nonvanishing, we need that $z \in J^-(y)$ and $y \in J^-(x)$ for all points for which $f(z) \neq 0$ and $g(x) \neq 0$. However, since the supports of $f$ and $g$ are spacelike separated, this is never the case, and the first integral vanishes. Analogously, one sees that also the second integral is vanishing. On the other hand, the last integral can be cancelled by a change of $f$
\begin{equation}
f(x) \to \hat{f}^\mathrm{adv}(x) = f(x) + \frac{\lambda}{6} ( G^\mathrm{adv} f )(x) \bigl[ ( \Delta f )(x) \bigr]^2 + \bigo{\lambda^2} \eqend{,}
\end{equation}
such that
\begin{equation}
\label{eq:phiret_fadv_support}
\int g(x) \varphi^\mathrm{ret}_{\hat{f}^\mathrm{adv}}(x) \total^d x = \int g(x) \left[ ( \Delta \hat{f}^\mathrm{adv} )(x) + \lambda \varphi_f^\mathrm{ret,(1)}(x) + \bigo{\lambda^2} \right] \total^d x = \bigo{\lambda^2} \eqend{.}
\end{equation}
It follows that $\supp \varphi^\mathrm{ret}_{\hat{f}^\mathrm{adv}} \subseteq J^+(\supp f) \cup J^-(\supp f)$, the same as for the free solution $\varphi_f^{(0)}$. Analogously, one shows that $\supp \varphi^\mathrm{adv}_{\hat{f}^\mathrm{ret}} \subseteq J^+(\supp f) \cup J^-(\supp f)$ with
\begin{equation}
\hat{f}^\mathrm{ret}(x) = f(x) + \frac{\lambda}{6} ( G^\mathrm{ret} f )(x) \bigl[ ( \Delta f )(x) \bigr]^2 + \bigo{\lambda^2} \eqend{,}
\end{equation}
and that $\supp \phiclass_{\bar{f}} \subseteq J^+(\supp f) \cup J^-(\supp f)$ with
\begin{equation}
\bar{f}(x) = f(x) + \frac{\lambda}{6} ( \overline{\Delta} f )(x) \bigl[ ( \Delta f )(x) \bigr]^2 + \bigo{\lambda^2} \eqend{.}
\end{equation}
Therefore, by suitably choosing $f$ one obtains a classical interacting solution with the same support properties as the free one.

Using that $\Delta = G^\mathrm{ret} - G^\mathrm{adv}$~\eqref{eq:delta_def} and that $\overline{\Delta} = \frac{1}{2} G^\mathrm{ret} + \frac{1}{2} G^\mathrm{adv}$~\eqref{eq:overlinedelta_def}, a short computa\-tion yields the explicit expression
\begin{splitequation}
\label{eq:phiclass_int_def}
\varphi^\mathrm{ret}_{\hat{f}^\mathrm{adv}}(x) &= \varphi^\mathrm{adv}_{\hat{f}^\mathrm{ret}}(x) = \phiclass_{\bar{f}} \equiv \varphi^\mathrm{int}_f \\
&= ( \Delta f )(x) + \frac{\lambda}{6} ( G^\mathrm{ret} \bigl[ ( G^\mathrm{ret} f ) ( \Delta f )^2 \bigr] )(x) - \frac{\lambda}{6} ( G^\mathrm{adv} \bigl[ ( G^\mathrm{adv} f ) ( \Delta f )^2 \bigr] )(x) + \bigo{\lambda^2} \eqend{.}
\end{splitequation}
That is, imposing causal evolution in the form of the support property $\supp \varphi^\mathrm{int}_f \subseteq J^+(\supp f) \cup J^-(\supp f)$, it seems that we obtain a unique classical interacting solution $\varphi^\mathrm{int}_f$. However, that is in fact not the case: since $\supp( ( G^\mathrm{ret} f ) ( G^\mathrm{adv} f ) ) \subseteq \supp( G^\mathrm{ret} f ) \cap \supp( G^\mathrm{adv} f ) = J^+(\supp f) \cap J^-(\supp f)$, it follows that
\begin{splitequation}
&\supp \Delta[ ( G^\mathrm{ret} f )^m ( G^\mathrm{adv} f )^n ] \\
&\quad\subseteq J^+\left( J^+(\supp f) \cap J^-(\supp f) \right) \cup J^-\left( J^+(\supp f) \cap J^-(\supp f) \right) \\
&\quad= J^+(\supp f) \cup J^-(\supp f)
\end{splitequation}
for arbitrary $m,n \geq 1$. Therefore, we may add arbitrary multiples of $\Delta[ ( G^\mathrm{ret} f ) ( G^\mathrm{adv} f )^2 ](x)$ and $\Delta[ ( G^\mathrm{ret} f )^2 ( G^\mathrm{adv} f ) ](x)$ to $\varphi^\mathrm{int}_f$, changing neither the support properties, nor the fact that it solves the non-linear Klein--Gordon equation up to linear order in $\lambda$, nor the invariance of the solution under the rescaling $f \to \alpha f$, $\lambda \to \alpha^{-2} \lambda$, $\varphi^\mathrm{int}_f \to \alpha \varphi^\mathrm{int}_f$.

\subsection{Classical Noether charges}
\label{sec:classical_noether}

Given a Killing vector $\xi^\mu$ of Minkowski spacetime, the associated classical Noether charge $Q_{\xi,x^0}$ is given by
\begin{equation}
Q_{\xi,x^0}(\varphi) = - \int_{x^0 = \text{const}} T_{0\mu}(\varphi) \xi^\mu \total^{d-1} \vec{x} \eqend{,}
\end{equation}
where $T_{\mu\nu}$ is the classical stress tensor~\eqref{eq:stress_tensor} specialized to Minkowski spacetime. As is well known, by Stokes' theorem one sees that $Q_{\xi,x^0}$ is independent of $x^0$ if $\varphi$ is space-compact and the stress tensor is conserved, which holds if $\varphi$ is a solution of the (interacting) equation of motion $E(\varphi) = 0$. The space-compactness condition holds, for example, for the perturbative solutions $\varphi^\mathrm{ret}_{\hat{f}^\mathrm{adv}}$, $\varphi^\mathrm{adv}_{\hat{f}^\mathrm{ret}}$ and $\phiclass_{\bar{f}}$ constructed in the last subsection, and we make this assumption from now on.

We are in particular interested in the translation charges $P^\alpha$ for which $\xi^\mu = \eta^{\mu\alpha}$ and the boost charges $M^{\alpha\beta}$ for which $\xi^\mu = - 2 \eta^{\mu[\alpha} x^{\beta]}$. We choose $x^0 = 0$ and obtain
\begin{equations}[eq:classical_noether_p]
P^0 &= \int_{x^0 = 0} T_{00}(\varphi) \total^{d-1} \vec{x} = \frac{1}{2} \int_{x^0 = 0} \left( \partial_0 \varphi \partial_0 \varphi + \partial^k \varphi \partial_k \varphi + m^2 \varphi^2 + \frac{\lambda}{12} \varphi^4 \right) \total^{d-1} \vec{x} \eqend{,} \\
P^i &= - \int_{x^0 = 0} T_0{}^i(\varphi) \total^{d-1} \vec{x} = - \int_{x^0 = 0} \partial_0 \varphi \partial^i \varphi \total^{d-1} \vec{x}
\end{equations}
for the translation charges and
\begin{equations}[eq:classical_noether_m]
M^{0i} &= - \int_{x^0 = 0} x^i T_{00}(\varphi) \total^{d-1} \vec{x} = - \frac{1}{2} \int_{x^0 = 0} x^i \left( \partial_0 \varphi \partial_0 \varphi + \partial^k \varphi \partial_k \varphi + m^2 \varphi^2 + \frac{\lambda}{12} \varphi^4 \right) \total^{d-1} \vec{x} \eqend{,} \\
M^{ij} &= - 2 \int_{x^0 = 0} x^{[i} T_0{}^{j]}(\varphi) \total^{d-1} \vec{x} = - 2 \int_{x^0 = 0} x^{[i} \partial^{j]} \varphi \partial_0 \varphi \total^{d-1} \vec{x}
\end{equations}
for the boost charges. We note that $H = P^0 \geq 0$, namely the classical energy is positive as long as $\lambda \geq 0$. Moreover, $M^{0i}(\varphi) \leq 0$ if $\supp \varphi \rvert_{x^0 = 0} \subseteq \mathcal{W}_\mathrm{R}$, a solution with support inside the Rindler wedge. Defining furthermore $K^i = - M^{0i}$ and $J_i = \frac{1}{2} \epsilon_{ijk} M^{jk}$ and using the classical Poisson bracket
\begin{equation}
\{ \varphi(\vec{x}), \partial_0 \varphi(\vec{y}) \} = - \{ \partial_0 \varphi(\vec{y}), \varphi(\vec{x}) \} = \delta^3(\vec{x}-\vec{y}) \eqend{,} \quad \{ \varphi(\vec{x}), \varphi(\vec{y}) \} = 0 = \{ \partial_0 \varphi(\vec{x}), \partial_0 \varphi(\vec{y}) \} \eqend{,}
\end{equation}
we compute that for space-compact $\varphi$ the Poincaré algebra
\begin{equations}[eq:classical_poincare_algebra_31]
\{ H, P^i \} &= 0 \eqend{,} \quad \{ H, J^i \} = 0 \eqend{,} \quad \{ H, K^i \} = - P^i \eqend{,} \\
\{ P^i, P^j \} &= 0 \eqend{,} \quad \{ J^i, P^j \} = \epsilon^{ijk} P_k \eqend{,} \quad \{ P^i, K^j \} = - \delta^{ij} H \eqend{,} \\
\{ J^i, K^j \} &= \epsilon^{ijk} K_k \eqend{,} \quad \{ J^i, J^j \} = \epsilon^{ijk} J_k \eqend{,} \quad \{ K^i, K^j \} = - \epsilon^{ijk} J_k
\end{equations}
is fulfilled.

\section{Quantization of \texorpdfstring{$\lambda \varphi^4$}{\textlambda\textphi\textfoursuperior} theory}
\label{sec:quantum}

We consider the quantized free scalar field $\phi(x)$, decomposed in creation and annihilation parts $\phi^+(x)$ and $\phi^-(x)$. We have the commutator $[ \phi(x), \phi(y) ] = \mathi \Delta(x,y)$ involving the commutator function~\eqref{eq:delta_def} and the two-point function
\begin{equation}
\label{eq:phi_2pf}
\bra{0} \phi(x) \phi(y) \ket{0} = \mathi G^+(x,y) = \mathi G^-(y,x)
\end{equation}
in the free Minkowski vacuum $\ket{0}$, from which it follows that
\begin{equation}
\label{eq:phi_commutator}
[ \phi^+(x), \phi^-(y) ] = - \mathi G^+(y,x) \eqend{,} \quad \Delta(x,y) = G^+(x,y) - G^+(y,x) \eqend{.}
\end{equation}
The commutator function fulfills the identities
\begin{equation}
\label{eq:delta_identities}
\Delta(x,y) \bigr\rvert_{x^0 = y^0} = 0 \eqend{,} \quad \partial_{x^0} \Delta(x,y) \bigr\rvert_{x^0 = y^0} = - \partial_{y^0} \Delta(x,y) \bigr\rvert_{x^0 = y^0} = - \delta^{d-1}(\vec{x}-\vec{y}) \eqend{,}
\end{equation}
which can be obtained directly from the explicit form~\eqref{eq:delta_def}, and which are equivalent to the canonical equal-time commutation relations, and the two-point function (and thus also the commutator) satisfies the Klein--Gordon equation
\begin{equation}
\left( \partial^2 - m^2 \right) G^+(x,y) = 0
\end{equation}
because the quantized field operator $\phi$ satisfies it.

The interacting quantized field $\Phi$ fulfills the normal-ordered interacting field equation
\begin{equation}
\left( \partial^2 - m^2 \right) \Phi = \frac{\lambda}{6} \normord{ \Phi^3 } \eqend{,}
\end{equation}
whose solution to first order is $\Phi = \phi + \lambda \Phi^{(1)} + \bigo{\lambda^2}$ with
\begin{equation}
\label{eq:phi1_correction}
\Phi^{(1)}(x) = \frac{1}{6} \int G^\mathrm{ret}(x,y) \normord{ \phi^3(y) } \total^d y \eqend{,}
\end{equation}
which corresponds to the retarded field and the Bogoliubov formula known from the algebraic approach to QFT~\cite{bogoliubovshirkov1959,duetschfredenhagen2001}. The interacting stress tensor $\mathcal{T}_{\mu\nu} = \mathcal{T}^{(0)}_{\mu\nu} + \lambda \mathcal{T}^{(1)}_{\mu\nu} + \bigo{\lambda^2}$ is also expanded in perturbation theory, and has two contributions: one from employing the interacting field, and one from the explicit interaction term. The free quantized stress tensor reads
\begin{equation}
\label{eq:stress_free}
\mathcal{T}^{(0)}_{\mu\nu} = \normord{ \partial_\mu \phi \partial_\nu \phi } - \frac{1}{2} \eta_{\mu\nu} \normord{ \partial^\rho \phi \partial_\rho \phi } - \frac{1}{2} m^2 \eta_{\mu\nu} \normord{ \phi^2 } \eqend{,}
\end{equation}
and the first-order correction is\footnote{The symmetrization is given by $A_{(\mu\nu)} = \frac{1}{2} A_{\mu\nu} + \frac{1}{2} A_{\nu\mu}$ for any tensor $A$.}
\begin{equation}
\label{eq:stress_order1_pre}
\mathcal{T}^{(1)}_{\mu\nu} = \left\{ \partial_{(\mu} \phi, \partial_{\nu)} \Phi^{(1)} \right\} - \frac{1}{2} \eta_{\mu\nu} \left\{ \partial^\rho \phi, \partial_\rho \Phi^{(1)} \right\} - \frac{1}{2} m^2 \eta_{\mu\nu} \left\{ \phi, \Phi^{(1)} \right\} - \frac{1}{4!} \eta_{\mu\nu} \normord{ \phi^4 } \eqend{.}
\end{equation}
While the free quantized stress tensor is already given as a well-defined operator on Fock space, to properly define the first-order correction we need to determine the symmetrized product $\left\{ \phi, \Phi^{(1)} \right\} = \phi \Phi^{(1)} + \Phi^{(1)} \phi$ and its derivatives. We therefore compute
\begin{splitequation}
\frac{1}{2} \left\{ \phi(x), \Phi^{(1)}(y) \right\} &= \frac{1}{6} \int G^\mathrm{ret}(y,z) \normord{ \phi(x) \phi^3(z) } \total^d z \\
&\quad+ \frac{\mathi}{4} \int \left[ G^+(x,z) + G^-(x,z) \right] G^\mathrm{ret}(y,z) \normord{ \phi^2(z) } \total^d z \eqend{,}
\end{splitequation}
and note that while the first integral is already a well-defined operator, the second one is divergent as $y \to x$, and so we must define a renormalized version of it.

In the coincidence limit $y \to x$, we obtain for $x \neq z$
\begin{splitequation}
\left[ G^+(x,z) + G^-(x,z) \right] G^\mathrm{ret}(x,z) &= \Theta(x^0-z^0) \left[ G^+(x,z) \right]^2 - \Theta(x^0-z^0) \left[ G^-(x,z) \right]^2 \\
&= \left[ G^\mathrm{F}(x,z) \right]^2 - \left[ G^-(x,z) \right]^2 \eqend{.}
\end{splitequation}
While the second term is well-defined as a distribution including for coinciding points $x = z$, the first one is divergent, and we need to replace it by a renormalized version. Taking the limit $y \to x$, we thus define
\begin{splitequation}
\label{eq:phi_phi1_def}
\left\{ \phi, \Phi^{(1)} \right\}(x) &= \frac{1}{3} \int G^\mathrm{ret}(x,z) \normord{ \phi(x) \phi^3(z) } \total^4 z \\
&\quad+ \frac{\mathi}{2} \int \left[ \left[ G^\mathrm{F}(x,z) \right]^2_\mathrm{ren} - \left[ G^-(x,z) \right]^2 \right] \normord{ \phi^2(z) } \total^4 z
\end{splitequation}
and
\begin{splitequation}
\label{eq:dphi_dphi1_def}
&\left\{ \partial_{(\mu} \phi, \partial_{\nu)} \Phi^{(1)} \right\}(x) = \frac{1}{3} \int \partial_{(\mu} G^\mathrm{ret}(x,z) \normord{ \partial_{\nu)} \phi(x) \phi^3(z) } \total^4 z \\
&\hspace{4em}+ \frac{\mathi}{2} \int \left[ \left[ \partial_\mu G^\mathrm{F}(x,z) \partial_\nu G^\mathrm{F}(x,z) \right]_\mathrm{ren} - \partial_\mu G^-(x,z) \partial_\nu G^-(x,z) \right] \normord{ \phi^2(z) } \total^4 z \eqend{,}
\end{splitequation}
where $\left[ G^\mathrm{F}(x,z) \right]^2_\mathrm{ren}$ and $\left[ \partial_\mu G^\mathrm{F}(x,z) \partial_\nu G^\mathrm{F}(x,z) \right]_\mathrm{ren}$ denote the renormalized distributions which we will determine explicitly later on. For the divergence of the so-defined stress tensor, we then obtain
\begin{equations}
\partial^\mu \mathcal{T}^{(0)}_{\mu\nu} &= \normord{ \left( \partial^2 - m^2 \right) \phi \partial_\nu \phi } = 0 \eqend{,} \\
\begin{split}
\partial^\mu \mathcal{T}^{(1)}_{\mu\nu} &= \frac{1}{6} \int \partial_\nu G^\mathrm{ret}(x,z) \normord{ \left( \partial^2 - m^2 \right) \phi(x) \phi^3(z) } \total^4 z \\
&\quad+ \frac{1}{6} \int \left( \partial^2 - m^2 \right) G^\mathrm{ret}(x,z) \normord{ \partial_\nu \phi(x) \phi^3(z) } \total^4 z \\
&\quad- \frac{\mathi}{2} \int \left( \partial^2 - m^2 \right) G^-(x,z) \partial_\nu G^-(x,z) \normord{ \phi^2(z) } \total^4 z - \frac{1}{6} \normord{ \partial_\nu \phi(x) \phi^3(x) } \\
&\quad+ \frac{\mathi}{4} \int \biggl[ 2 \partial^\mu \left[ \partial_\mu G^\mathrm{F}(x,z) \partial_\nu G^\mathrm{F}(x,z) \right]_\mathrm{ren} - \partial_\nu \left[ \partial_\mu G^\mathrm{F}(x,z) \partial^\mu G^\mathrm{F}(x,z) \right]_\mathrm{ren} \\
&\hspace{4em}- m^2 \partial_\nu \left[ G^\mathrm{F}(x,z) \right]^2_\mathrm{ren} \biggr] \normord{ \phi^2(z) } \total^4 z \\
&= \frac{\mathi}{4} \int \biggl[ 2 \partial^\mu \left[ \partial_\mu G^\mathrm{F}(x,z) \partial_\nu G^\mathrm{F}(x,z) \right]_\mathrm{ren} - \partial_\nu \left[ \partial_\mu G^\mathrm{F}(x,z) \partial^\mu G^\mathrm{F}(x,z) \right]_\mathrm{ren} \\
&\hspace{4em}- m^2 \partial_\nu \left[ G^\mathrm{F}(x,z) \right]^2_\mathrm{ren} \biggr] \normord{ \phi^2(z) } \total^4 z \eqend{,}
\end{split}
\end{equations}
where we used that $\phi$ and $G^-$ fulfill the free Klein--Gordon equation and that
\begin{equation}
\left( \partial^2 - m^2 \right) G^\mathrm{ret}(x,z) = \delta(x,z) \eqend{.}
\end{equation}
To obtain a conserved stress tensor also in the quantum theory, we thus need to define our renormalization in such a way that the last line vanishes:
\begin{equation}
\label{eq:stress_conserved_condition}
2 \partial^\mu \left[ \partial_\mu G^\mathrm{F}(x,y) \partial_\nu G^\mathrm{F}(x,y) \right]_\mathrm{ren} - \partial_\nu \left[ \partial_\mu G^\mathrm{F}(x,y) \partial^\mu G^\mathrm{F}(x,y) \right]_\mathrm{ren} - m^2 \partial_\nu \left[ G^\mathrm{F}(x,y) \right]^2_\mathrm{ren} = 0 \eqend{.}
\end{equation}
This can be achieved in general if one uses Hadamard normal ordering~\cite{hollandswald2005}, but we also show it explicitly in appendix~\ref{app:renormalization}. The results are given in equations~\eqref{eq:app_gfren2} and~\eqref{eq:app_dgfren2}.

Inserting the renormalized results~\eqref{eq:phi_phi1_def} and~\eqref{eq:dphi_dphi1_def} into the stress tensor correction~\eqref{eq:stress_order1_pre} and using the equations~\eqref{eq:app_gf_stress} and~\eqref{eq:app_gpm_stress} from appendix~\ref{app:renormalization}, we obtain the renormalized first-order correction
\begin{splitequation}
\label{eq:stress_order1}
\mathcal{T}^{(1),\mathrm{ren}}_{\mu\nu} &= \frac{1}{3} \left( \delta_\mu^\alpha \delta_\nu^\beta - \frac{1}{2} \eta_{\mu\nu} \eta^{\alpha\beta} \right) \int \partial_{(\alpha} G^\mathrm{ret}(x,z) \normord{ \partial_{\beta)} \phi(x) \phi^3(z) } \total^4 z \\
&\quad- \frac{1}{6} m^2 \eta_{\mu\nu} \int G^\mathrm{ret}(x,z) \normord{ \phi(x) \phi^3(z) } \total^4 z - \frac{1}{4!} \eta_{\mu\nu} \normord{ \phi^4 } \\
&\quad+ \frac{1}{6 (4 \pi)^2} \left( \partial_\mu \partial_\nu - \eta_{\mu\nu} \partial^2 \right) \int \biggl[ I^\mathrm{F}_0(x,z) - 2 I^\mathrm{F}_1(x,z) - 2 m^2 G^\mathrm{F}_0(x,z) \\
&\hspace{6em}- I^-_0(x,z) + 2 I^-_1(x,z) + 6 \mathi (4 \pi)^2 c' \delta^4(x-z) \biggr] \normord{ \phi^2(z) } \total^4 z \eqend{,}
\end{splitequation}
where $c'$ is an arbitrary constant and the integrals $I^\mathrm{F}_k$ and $I^-_k$ are defined by equations~\eqref{eq:app_ifk_def} and~\eqref{eq:app_ipk_def}, the latter with the substitution $\Theta(p^0) \to \Theta(-p^0)$.

\subsection{Quantum Noether charges}
\label{sec:quantum_noether}

Having a well-defined quantized interacting stress tensor at hand, we proceed to define the Noether charges in the quantum theory. For this, we first consider the commutator of the stress tensor and the interacting field $\Phi$.

In the free theory, we compute straightforwardly that
\begin{equation}
- \mathi \left[ \mathcal{T}^{(0)}_{\mu\nu}(x), \phi(y) \right] = 2 \partial_{(\mu} \phi(x) \partial_{\nu)} \Delta(x,y) - \eta_{\mu\nu} \partial^\rho \phi(x) \partial_\rho \Delta(x,y) - m^2 \eta_{\mu\nu} \phi(x) \Delta(x,y) \eqend{,}
\end{equation}
where all derivatives act on $x$. Restricting to $x^0 = y^0$ and contracting with the Killing vector $\xi^\nu$, we obtain
\begin{splitequation}
- \mathi \left[ \xi^\nu \mathcal{T}^{(0)}_{0\nu}(x), \phi(y) \right]_{x^0 = y^0} &= - 2 \xi^{[k} \partial^{0]} \phi(x) \partial_k \Delta(x,y) \bigr\rvert_{x^0 = y^0} \\
&\quad+ \xi^\mu \partial_\mu \phi(x) \partial_0 \Delta(x,y) \bigr\rvert_{x^0 = y^0} + m^2 \xi^0 \phi(x) \Delta(x,y) \bigr\rvert_{x^0 = y^0} \eqend{.}
\end{splitequation}
We may now employ the identities~\eqref{eq:delta_identities} of the commutator function to obtain
\begin{equation}
\mathi \left[ \xi^\nu \mathcal{T}^{(0)}_{0\nu}(x), \phi(y) \right]_{x^0 = y^0} = \xi^\mu \partial_\mu \phi(x) \delta^3(\vec{x}-\vec{y}) \eqend{,}
\end{equation}
and integrating with any function of $\vec{x}$ that is equal to $1$ in a neighborhood of $\vec{y}$, it follows that
\begin{equation}
\label{eq:free_commutator_sigma}
\left[ \mathcal{Q}_{\xi,f}, \phi(y) \right] = \mathi \xi^\mu \partial_\mu \phi(y)
\end{equation}
with the charge
\begin{equation}
\label{eq:free_charge_stress_tensor}
\mathcal{Q}_{\xi,f} = - \int_{x^0 = \text{const}} f(\vec{x}) \xi^\nu \mathcal{T}^{(0)}_{0\nu}(x) \total^3 \vec{x} \eqend{.}
\end{equation}
Since we have ensured that $\partial^\mu \mathcal{T}^{(0)}_{\mu\nu} = 0$, as in the classical case the quantum Noether charge $\mathcal{Q}_{\xi,f}$ is independent of $x^0$. We may thus also consider a function $f(x) = f_\mathrm{t}(x^0) f_\mathrm{s}(\vec{x})$ with $\int f_\mathrm{t}(x^0) \total x^0 = 1$ and $f_\mathrm{s}$ equal to $1$ in a neighborhood of $\vec{y}$. Clearly the commutator~\eqref{eq:free_commutator_sigma} is still valid for this choice of $f$, and the commutator on $x^0 = \text{const}$ is recovered in the limit $f_\mathrm{t}(x^0) \to \delta(x^0)$.

In the interacting theory, using the explicit expressions~\eqref{eq:phi_phi1_def} and~\eqref{eq:dphi_dphi1_def} we compute the commutators
\begin{splitequation}
\left[ \left\{ \phi, \Phi^{(1)} \right\}(x), \phi(y) \right] &= 2 \mathi \Delta(x,y) \Phi^{(1)}(x) - \mathi \int G^\mathrm{ret}(x,z) \Delta(y,z) \normord{ \phi(x) \phi^2(z) } \total^4 z \\
&\quad+ \int \left[ \left[ G^\mathrm{F}(x,z) \right]^2_\mathrm{ren} - \left[ G^-(x,z) \right]^2 \right] \Delta(y,z) \phi(z) \total^4 z
\end{splitequation}
and
\begin{splitequation}
&\left[ \left\{ \partial_{(\mu} \phi, \partial_{\nu)} \Phi^{(1)} \right\}(x), \phi(y) \right] \\
&\quad= 2 \mathi \partial_{(\mu} \Delta(x,y) \partial_{\nu)} \Phi^{(1)}(x) - \mathi \int \Delta(y,z) \partial_{(\mu} G^\mathrm{ret}(x,z) \normord{ \partial_{\nu)} \phi(x) \phi^2(z) } \total^4 z \\
&\qquad+ \int \left[ \left[ \partial_\mu G^\mathrm{F}(x,z) \partial_\nu G^\mathrm{F}(x,z) \right]_\mathrm{ren} - \partial_\mu G^-(x,z) \partial_\nu G^-(x,z) \right] \Delta(y,z) \phi(z) \total^4 z \eqend{,}
\end{splitequation}
as well as
\begin{equation}
[ \Phi^{(1)}(x), \phi(y) ] = - \frac{\mathi}{2} \int G^\mathrm{ret}(x,z) \Delta(y,z) \normord{ \phi^2(z) } \total^4 z \eqend{.}
\end{equation}
A long but straightforward computation then results in
\begin{splitequation}
- \mathi \left[ \mathcal{T}_{\mu\nu}(x), \Phi(y) \right] &= 2 \partial_{(\mu} \Phi(x) \partial_{\nu)} \Delta(x,y) - \eta_{\mu\nu} \partial^\rho \Phi(x) \partial_\rho \Delta(x,y) - m^2 \eta_{\mu\nu} \Phi(x) \Delta(x,y) \\
&\quad+ \lambda \biggl[ \int \left[ G^\mathrm{ret}(y,z) \partial_{(\mu} \Delta(x,z) - \Delta(y,z) \partial_{(\mu} G^\mathrm{ret}(x,z) \right] \normord{ \partial_{\nu)} \phi(x) \phi^2(z) } \total^4 z \\
&\qquad\quad- \frac{1}{2} \eta_{\mu\nu} \int \left[ G^\mathrm{ret}(y,z) \partial^\rho \Delta(x,z) - \Delta(y,z) \partial^\rho G^\mathrm{ret}(x,z) \right] \normord{ \partial_\rho \phi(x) \phi^2(z) } \total^4 z \\
&\qquad\quad+ \frac{1}{2} \eta_{\mu\nu} m^2 \int \left[ G^\mathrm{ret}(x,z) \Delta(y,z) - \Delta(x,z) G^\mathrm{ret}(y,z) \right] \normord{ \phi(x) \phi^2(z) } \total^4 z \\
&\qquad\quad- \frac{\mathi}{2} \int \left[ H^+_{\mu\nu}(x,z) G^\mathrm{ret}(y,z) - H^-_{\mu\nu}(x,z) G^\mathrm{adv}(y,z) \right] \phi(z) \total^4 z \\
&\qquad\quad- \frac{1}{6} \eta_{\mu\nu} \Delta(x,y) \normord{ \phi^3(x) } \biggr] + \bigo{ \lambda^2 }
\end{splitequation}
with the combination
\begin{splitequation}
H^\pm_{\mu\nu}(x,y,z) &= 2 \left[ \partial_{(\mu} G^\mathrm{F}(x,z) \partial_{\nu)} G^\mathrm{F}(x,z) \right]_\mathrm{ren} - 2 \partial_{(\mu} G^\pm(x,z) \partial_{\nu)} G^\pm(x,z) \\
&\quad- \eta_{\mu\nu} \left[ \left[ \partial_\rho G^\mathrm{F}(x,z) \partial^\rho G^\mathrm{F}(x,z) \right]_\mathrm{ren} - \partial_\rho G^\pm(x,z) \partial^\rho G^\pm(x,z) \right] \\
&\quad- \eta_{\mu\nu} m^2 \left[ \left[ G^\mathrm{F}(x,z) \right]^2_\mathrm{ren} - \left[ G^\pm(x,z) \right]^2 \right] \eqend{,}
\end{splitequation}
where all derivatives act on $x$, and where we used that
\begin{equation}
\phi(x) \normord{ \phi^2(z) } + \normord{ \phi^2(z) } \phi(x) = 2 \normord{ \phi(x) \phi^2(z) } + 2 \mathi [ G^+(x,z) + G^-(x,z) ] \phi(z) \eqend{,}
\end{equation}
and that
\begin{equations}
\Delta(x,y) \left[ G^+(x,y) + G^-(x,y) \right] &= \left[ G^+(x,y) \right]^2 - \left[ G^-(x,y) \right]^2 \eqend{,} \\
\partial_{(\mu} \Delta(x,y) \partial_{\nu)} \left[ G^+(x,y) + G^-(x,y) \right] &= \partial_\mu G^+(x,y) \partial_\nu G^+(x,y) - \partial_\mu G^-(x,y) \partial_\nu G^-(x,y) \eqend{.}
\end{equations}

As before, it follows straightforwardly that the terms in the first line give the required result when integrated against a function $f$ of the form $f(x) = f_\mathrm{t}(x^0) f_\mathrm{s}(\vec{x})$ as explained above, and we thus have only to show that the terms proportional to $\lambda$ vanish. For the first three integrals and the term proportional to $\normord{ \phi^3(x) }$ this follows easily, restricting again to $x^0 = y^0$ (i.e., considering the limit $f_\mathrm{t}(x^0) \to \delta(x^0)$) and using that
\begin{equations}
\left[ G^\mathrm{ret}(y,z) \Delta(x,z) - \Delta(y,z) G^\mathrm{ret}(x,z) \right]_{x^0 = y^0} &= 0 \eqend{,} \\
\left[ G^\mathrm{ret}(y,z) \partial_\mu \Delta(x,z) - \Delta(y,z) \partial_\mu G^\mathrm{ret}(x,z) \right]_{x^0 = y^0} &= 0 \eqend{.}
\end{equations}
For the integral involving $H^\pm_{\mu\nu}$, we have to work a bit more. We first note that by our normalization condition~\eqref{eq:stress_conserved_condition} that we imposed to obtain a conserved stress tensor, the terms in $\partial^\mu H^\pm_{\mu\nu}$ involving products of the Feynman propagator vanish, while the terms involving the two-point functions $G^\pm$ vanish by direct computation, using that $G^\pm$ are solutions of the free Klein--Gordon equation, such that $\partial^\mu H^\pm_{\mu\nu} = 0$. Moreover, from the definition of the Feynman propagator we see that $H^\pm_{\mu\nu}(x,z) = 0$ if $z \neq J^\pm(x)$, i.e., $H^+_{\mu\nu}(x,z)$ has the same support as the advanced propagator $G^\mathrm{adv}(x,z)$ and $H^-_{\mu\nu}(x,z)$ has the same support as the retarded propagator $G^\mathrm{ret}(x,z)$.

Assume now that we can decompose
\begin{equation}
\label{eq:f_decomposition}
f \delta^\mu_0 = f_\pm^\mu + \partial^\mu \hat{f}_\pm \eqend{,}
\end{equation}
where all functions are compactly supported and $\supp f_\pm^\mu \cap J^\mp(y) = \emptyset$. Integrating by parts, we obtain
\begin{splitequation}
&\int f(x) \xi^\nu(x) \int H^+_{0\nu}(x,z) G^\mathrm{ret}(y,z) \phi(z) \total^4 z \total^4 x \\
&\quad= \iint f_+^\mu(x) \xi^\nu(x) H^+_{\mu\nu}(x,z) G^\mathrm{ret}(y,z) \phi(z) \total^4 z \total^4 x \\
&\qquad- \iint \hat{f}_+(x) \partial^\mu \left[ \xi^\nu(x) H^+_{\mu\nu}(x,z) \right] G^\mathrm{ret}(y,z) \phi(z) \total^4 z \total^4 x \eqend{,}
\end{splitequation}
and since $H^+_{\mu\nu}$ is conserved and symmetric and $\partial^{(\mu} \xi^{\nu)} = 0$ since $\xi^\mu$ is a Killing vector, the integral in the last line vanishes. On the other hand, to obtain a nonvanishing result for the first integral we need $z \in J^-(y)$ by the support properties of the retarded propagator and $x \in J^-(z)$ by the support properties of $H^+_{\mu\nu}(x,z)$, and hence $x \in J^-(y)$. However, then $f_+^\mu(x)$ vanishes, and so does the whole integral. The analogous argument shows that also the integral containing $H^-_{\mu\nu}$ vanishes, and it remains to show the decomposition~\eqref{eq:f_decomposition}.

Since $f(x) = f_\mathrm{t}(x^0) f_\mathrm{s}(\vec{x})$ is equal to $1$ in a neighborhood of $y$, it is no loss of generality to assume that $\supp f_\mathrm{t}$ is small enough such that there exists a neighborhood of the past light cone of $y$ where $f_\mathrm{s}(\vec{x}) = 1$. Therefore, $f(x)$ restricted to this neighborhood only depends on time, and we can clearly write this restriction as $\partial_{x^0} \hat{f}_+$ for some function $\hat{f}_+$, which is compactly supported and only depends on $x^0$ inside this neighborhood. The remainder $f^\mu_+ = f(x) n^\mu - \partial^\mu \hat{f}_+$ therefore vanishes in this neighborhood, and we obtain the required decomposition. Replacing $J^-$ by $J^+$, the analogous arguments also show the decomposition into $f^\mu_-$ and $\hat{f}_-$ with the required support properties, such that~\eqref{eq:f_decomposition} holds.

A generalization of this proof to curved spacetimes and to arbitrary order in the interaction can be found in~\cite{hollands2001}, and we may thus define the interacting quantum Noether charges
\begin{equation}
\mathcal{Q}_{\xi,f} = - \int f(x) \xi^\nu \mathcal{T}_{0\nu}(x) \total^4 x \eqend{,}
\end{equation}
which give the correct commutators
\begin{equation}
\label{eq:commutator_charge_field}
[ \mathcal{Q}_{\xi,f}, \Phi(x) ] = \mathi \xi^\mu \partial_\mu \Phi(x)
\end{equation}
for all functions $f$ with the above properties. In particular, choosing $\xi^\rho = \eta^{\mu\rho}$ and $\xi^\rho = - 2 \eta^{\rho[\mu} x^{\nu]}$ we obtain the generators of Lorentz transformations in the quantum theory
\begin{equations}[eq:minkowski_lorentz_generators]
\mathcal{P}^\mu(f) &= \int f(x) \mathcal{T}^{0\mu}(x) \total^4 x \eqend{,} \\
\mathcal{M}^{\mu\nu}(f) &= \int f(x) \left[ x^\mu \mathcal{T}^{0\nu}(x) - x^\nu \mathcal{T}^{0\mu}(x) \right] \total^4 x \eqend{,}
\end{equations}
which satisfy the Poincaré algebra in the adiabatic limit $f(x) \to \delta(x^0)$. While this limit might not exist for the charges themselves, it always exists for their commutators~\cite{requardt1976}. However, since the commutator is well-defined and actually independent of $f$ as long as $f = 1$ in a neighborhood of $y$, we may drop the dependence on $f$ in the notation of the charges, and in the following write $\mathcal{Q}_\xi$ for them.

\subsection{Interacting Weyl algebra}
\label{sec:quantum_weyl}

Having defined the interacting field operator, stress tensor and Noether charges, we proceed to construct the von Neumann algebra $\mathcal{A}$ of the Rindler wedge $\mathcal{W}_\mathrm{R}$ in the interacting theory. This algebra is generated by interacting Weyl operators
\begin{equation}
\label{eq:weyl_def}
W(f) = \mathe^{\mathi \Phi(f)} \eqend{,}
\end{equation}
where $\Phi = \phi + \lambda \Phi^{(1)} + \bigo{\lambda^2}$ is the interacting field operator whose first-order correction is given by~\eqref{eq:phi1_correction}. Smearing with a test function $f$, this explicitly reads
\begin{splitequation}
\label{eq:phi_int_f}
\Phi(f) &= \phi(f) + \frac{\lambda}{6} \iint f(x) G^\mathrm{ret}(x,y) \normord{ \phi^3(y) } \total^4 y + \bigo{\lambda^2} \\
&= \phi(f) + \frac{\lambda}{6} \int ( G^\mathrm{adv} f )(y) \normord{ \phi^3(y) } \total^4 y + \bigo{\lambda^2} \eqend{.}
\end{splitequation}
Using the formula
\begin{equation}
\partial_t \, \mathe^{X(t)} \Bigr\rvert_{t = 0} = \mathe^{X(0)} \frac{1 - \mathe^{- \operatorname{ad}_{X(0)}}}{\operatorname{ad}_{X(0)}} \partial_t X(t) \Bigr\rvert_{t = 0}
\end{equation}
for the derivative of the exponential map at 0, we also obtain
\begin{splitequation}
\label{eq:weyl_expansion_r}
W(f) &= \mathe^{\mathi \Phi(f)} \Bigr\rvert_{\lambda = 0} + \lambda \left[ \partial_\lambda \mathe^{\mathi \Phi(f)} \right]_{\lambda = 0} + \bigo{\lambda^2} \\
&= \mathe^{\mathi \phi(f)} \biggl[ 1 + \frac{\mathi \lambda}{6} \int ( G^\mathrm{adv} f )(z) \normord{ \phi^3(z) } \total^4 z - \frac{\mathi \lambda}{4} \int ( G^\mathrm{adv} f )(z) ( \Delta f )(z) \normord{ \phi^2(z) } \total^4 z \\
&\qquad+ \frac{\mathi \lambda}{6} \int ( G^\mathrm{adv} f )(z) [ ( \Delta f )(z) ]^2 \phi(z) \total^4 z - \frac{\mathi \lambda}{24} \int ( G^\mathrm{adv} f )(z) [ ( \Delta f )(z) ]^3 \total^4 z + \bigo{\lambda^2} \biggr] \eqend{.}
\end{splitequation}
Commuting fields to the left, we also have
\begin{splitequation}
\label{eq:weyl_expansion_l}
W(f) &= \biggl[ 1 + \frac{\mathi \lambda}{6} \int ( G^\mathrm{adv} f )(z) \normord{ \phi^3(z) } \total^4 z + \frac{\mathi \lambda}{4} \int ( G^\mathrm{adv} f )(z) ( \Delta f )(z) \normord{ \phi^2(z) } \total^4 z \\
&\quad+ \frac{\mathi \lambda}{6} \int ( G^\mathrm{adv} f )(z) [ ( \Delta f )(z) ]^2 \phi(z) \total^4 z + \frac{\mathi \lambda}{24} \int ( G^\mathrm{adv} f )(z) [ ( \Delta f )(z) ]^3 \total^4 z + \bigo{\lambda^2} \biggr] \mathe^{\mathi \phi(f)} \eqend{,}
\end{splitequation}
where we employed the commutators
\begin{equations}[eq:weyl_commutators]
\left[ \mathe^{\mathi \phi(f)}, \phi(x) \right] &= \mathe^{\mathi \phi(f)} ( \Delta f )(x) \eqend{,} \\
\left[ \mathe^{\mathi \phi(f)}, \normord{ \phi^2(x) } \right] &= \mathe^{\mathi \phi(f)} \left[ 2 ( \Delta f )(x) \phi(x) - [ ( \Delta f )(x) ]^2 \right] \eqend{,} \\
\left[ \mathe^{\mathi \phi(f)}, \normord{ \phi^3(x) } \right] &= \mathe^{\mathi \phi(f)} \left[ 3 ( \Delta f )(x) \normord{ \phi^2(x) } - 3 [ ( \Delta f )(x) ]^2 \phi(x) + [ ( \Delta f )(x) ]^3 \right] \eqend{.}
\end{equations}
From these formulas, it is easy to see that $[ W(f) ]^\dagger = W(-f)$ for real $f$, using that the free quantum field is Hermitean: $\phi^\dagger = \phi$. Since also the interacting field~\eqref{eq:phi_int_f} is Hermitean for real $f$, the interacting Weyl operators are bounded, and to define the von Neumann algebra $\mathcal{A}$ of the Rindler wedge $\mathcal{W}_\mathrm{R}$, it only remains to verify their support properties.

For this, we first compute the product of two interacting Weyl operators
\begin{splitequation}
\label{eq:weyl_product}
W(f) W(g) &= W(f+g) \, \mathe^{- \frac{\mathi}{2} \Delta(f,g)} \biggl[ 1 - \frac{\mathi \lambda}{24} \int \left[ ( G^\mathrm{adv} f )(z) ( \Delta g )(z) - ( G^\mathrm{adv} g )(z) ( \Delta f )(z) \right] \\
&\hspace{5em}\times \Bigl[ 6 \normord{ \phi^2(z) } - 4 \left[ ( \Delta f )(z) + 2 ( \Delta g )(z) \right] \phi(z) \\
&\hspace{6em}+ [ ( \Delta f )(z) ]^2 + 3 [ ( \Delta g )(z) ]^2 + 3 ( \Delta f )(z) ( \Delta g )(z) \Bigr] \total^4 z + \bigo{\lambda^2} \biggr] \eqend{,}
\end{splitequation}
where we used the expansion~\eqref{eq:weyl_expansion_r} for the interacting Weyl operators, the commutators~\eqref{eq:weyl_commutators} to bring all the fields to the right, and the special form
\begin{equation}
\mathe^{\mathi \phi(f)} \mathe^{\mathi \phi(g)} = \mathe^{\mathi \phi(f+g)} \mathe^{- \frac{\mathi}{2} \Delta(f,g)}
\end{equation}
of the Baker--Campbell--Hausdorff formula which is valid if the commutator is proportional to the identity. Here, we defined
\begin{equation}
\label{eq:delta_fg_def}
\Delta(f,g) = \iint f(x) \Delta(x,y) g(y) \total^4 x \total^4 y = \int f(x) ( \Delta g )(x) \total^4 x \eqend{.}
\end{equation}
Consider then $f$ and $g$ whose support is spacelike separated. Since $\supp( \Delta g ) \subseteq J^+( \supp g ) \cup J^-( \supp g)$, it follows that $\Delta g$ vanishes on the support of $f$ such that $\Delta(f,g) = 0$. Moreover, since
\begin{splitequation}
&( G^\mathrm{adv} f )(z) ( \Delta g )(z) - ( G^\mathrm{adv} g )(z) ( \Delta f )(z) \\
&= - \iint f(x) g(y) \Delta(x,z) \Delta(y,z) \left[ \Theta(x^0-z^0) - \Theta(y^0-z^0) \right] \total^4 x \total^4 y \eqend{,}
\end{splitequation}
we see that $z \in J^-(x) \cap J^+(y)$ must hold for this expression to be nonzero. However, if $f$ and $g$ are spacelike separated, these cones do not intersect, and so it vanishes. It follows that $W(f) W(g) = W(f+g) = W(g) W(f)$ if $f$ and $g$ are spacelike separated, and therefore the interacting Weyl operators commute in this case. The same holds for the commutator of two interacting fields:
\begin{splitequation}
[ \Phi(f), \Phi(g) ] &= \mathi \Delta(f,g) \\
&\quad+ \frac{\mathi \lambda}{2} \int \left[ ( G^\mathrm{adv} f )(z) ( \Delta g )(z) - ( G^\mathrm{adv} g )(z) ( \Delta f )(z) \right] \normord{ \phi^2(z) } \total^4 z + \bigo{\lambda^2} \eqend{.}
\end{splitequation}
We may thus define the von Neumann algebra $\mathcal{A}$ completely analogously to the free theory, namely as the double commutant of the span of interacting Weyl operators with functions whose support lies in the Rindler wedge:
\begin{equation}
\mathcal{A} = \left\{ \sum_i c_i W(f_i) \colon c_i \in \mathbb{C}, \supp f_i \subset \mathcal{W}_\mathrm{R} \right\}'' \eqend{.}
\end{equation}
Because $W(f)$ and $W(g)$ commute if the supports of $f$ and $g$ are spacelike separated, the commutant algebra $\mathcal{A}'$ is generated by interacting Weyl operators with functions whose support lies in the left Rindler wedge $\mathcal{W}_\mathrm{L} = \{ x \in \mathbb{R}^{1,d-1} \colon x^1 \leq - \abs{x^0} \}$. Haag duality is thus fulfilled, and the Bisognano--Wichmann theorem is applicable.

Exponentiating the commutator~\eqref{eq:commutator_charge_field} of the interacting quantum Noether charge and the interacting field, it follows that
\begin{equation}
\mathe^{\mathi \mathcal{Q}_\xi} \Phi(x) \mathe^{- \mathi \mathcal{Q}_\xi} = \mathe^{- \xi^\mu \partial_\mu} \Phi(x) \eqend{,}
\end{equation}
and thus on smeared fields
\begin{equation}
\label{eq:qxi_phi}
\mathe^{\mathi \mathcal{Q}_\xi} \Phi(f) \mathe^{- \mathi \mathcal{Q}_\xi} = \Phi(f_\xi) \quad\text{with}\quad f_\xi(x) = \mathe^{\xi^\mu \partial_\mu} f(x) \eqend{.}
\end{equation}
The same action then holds on the interacting Weyl operators:
\begin{equation}
\label{eq:qxi_weyl}
\mathe^{\mathi \mathcal{Q}_\xi} W(f) \mathe^{- \mathi \mathcal{Q}_\xi} = W(f_\xi) \eqend{.}
\end{equation}

Finally, for the computation of the relative entropy we need the vacuum expectation value of an interacting Weyl operator and a normal-ordered product of free fields. We recall that a free Weyl operator $\mathe^{\mathi \phi(f)}$ has the expectation value
\begin{equation}
\bra{0} \mathe^{\mathi \phi(f)} \ket{0} = \mathe^{- \frac{\mathi}{2} G^+(f,f)}
\end{equation}
in the free Minkowski vacuum $\ket{0}$, which can be computed by expanding the exponential and using Wick's theorem together with the two-point function~\eqref{eq:phi_2pf}. Together with a normal-ordered product of free fields, we obtain analogously
\begin{equation}
\label{eq:free_weyl_normalordered_expectation}
\bra{0} \mathe^{\mathi \phi(f)} \normord{ \phi(x_1) \cdots \phi(x_k) } \ket{0} = \mathe^{- \frac{\mathi}{2} G^+(f,f)} \prod_{j=1}^k [ - (G^- f)(x_j) ] \eqend{.}
\end{equation}
Using then the expansion~\eqref{eq:weyl_expansion_r} for the interacting Weyl operators and the rule
\begin{equation}
\normord{ A(\phi) } \, \normord{ B(\phi) } = \normord{ A(\phi) \exp\left( \mathi \iint \overleftarrow{\frac{\delta}{\delta \phi(x)}} G^+(x,y) \overrightarrow{\frac{\delta}{\delta \phi(y)}} \total^4 x \total^4 y \right) B(\phi) }
\end{equation}
for the product of two normal-ordered operators $A(\phi)$ and $B(\phi)$ depending on the free field $\phi$, a long but straightforward computation results in
\begin{splitequation}
\label{eq:weyl_normalordered_expectation}
&\bra{0} W(f) \normord{ \phi(x_1) \cdots \phi(x_k) } \ket{0} = \mathe^{- \frac{\mathi}{2} G^+(f,f)} \prod_{j=1}^k [ - (G^- f)(x_j) ] \\
&\quad\times \biggl[ 1 - \frac{\mathi \lambda}{24} \int ( G^\mathrm{adv} f )(z) \biggl[ \left[ (G^- f)(z) + (G^+ f)(z) \right] \left[ [ (G^- f)(z) ]^2 + [ (G^+ f)(z) ]^2 \right] \\
&\qquad\qquad+ 4 \mathi \left[ [ (G^- f)(z) ]^2 + [ (G^+ f)(z) ]^2 + (G^- f)(z) (G^+ f)(z) \right] \sum_{j=1}^k \frac{G^-(x_j,z)}{(G^- f)(x_j)} \\
&\qquad\qquad- 12 \left[ (G^- f)(z) + (G^+ f)(z) \right] \sum_{1 \leq j < j' \leq k} \frac{G^-(x_j,z) G^-(x_{j'},z)}{(G^- f)(x_j) (G^- f)(x_{j'})} \\
&\qquad\qquad- 24 \mathi \sum_{1 \leq j < j' < j'' \leq k} \frac{G^-(x_j,z) G^-(x_{j'},z) G^-(x_{j''},z)}{(G^- f)(x_j) (G^- f)(x_{j'}) (G^- f)(x_{j''})} \biggr] \total^4 z + \bigo{\lambda^2} \biggr] \eqend{.}
\end{splitequation}

\subsection{Interacting vacuum state}

As a final step in the construction, we also need to determine the interacting vacuum state $\ket{\Omega}$. This also has a perturbative expansion
\begin{equation}
\label{eq:interacting_omega_expansion}
\ket{\Omega} = \ket{0} + \lambda \ket[\big]{ \Omega^{(1)} } + \bigo{\lambda^2} \eqend{,}
\end{equation}
and we are interested in the first-order correction $\ket[\big]{ \Omega^{(1)} }$. In both time-dependent (Feynman--Dyson) and time-independent (Rayleigh--Schrödinger) perturbation theory, one expands $\ket[\big]{ \Omega^{(1)} }$ in a basis of free-theory Fock states $\ket{\vec{n}_1,\ldots,\vec{n}_k}$ for $k \in \mathbb{N}_0$, where each label $\vec{n}_j$ describes an element in a basis of the single-particle Hilbert space. Starting with the free Minkowski vacuum $\ket{0}$ in the free theory, the expansion coefficient of $\ket{\vec{n}_1,\ldots,\vec{n}_k}$ in $\ket[\big]{ \Omega^{(1)} }$ is proportional to the matrix element $\bra{\vec{n}_1,\ldots,\vec{n}_k} V \ket{0}$, where $V$ is the perturbation to the Hamiltonian. Therefore, choosing a basis of the form $\ket{\vec{n}_1,\ldots,\vec{n}_k} \sim \normord{ \phi(g_{\vec{n}_1}) \cdots \phi(g_{\vec{n}_k}) } \ket{0}$ with suitable functions $g_\vec{j}$ and using Wick's theorem, we see that for the normal-ordered $\phi^4$ interaction only the coefficient with $k = 4$ is different from zero. Hence, $\ket[\big]{ \Omega^{(1)} }$ will be proportional to a sum of terms $\normord{ \phi(g_{\vec{n}_1}) \cdots \phi(g_{\vec{n}_k}) } \ket{0}$ with $k = 4$.

For the computation of the relative entropy, we need to compute the matrix elements $\bra[\big]{0} \mathe^{\mathi \phi(f)} \ket[\big]{ \Omega^{(1)} }$ and $\bra[\big]{ \Omega^{(1)} } \mathe^{\mathi \phi(f)} \ket[\big]{0}$ of a free Weyl operator $\mathe^{\mathi \phi(f)}$. Using Wick's theorem and the result~\eqref{eq:free_weyl_normalordered_expectation}, it thus follows that
\begin{equation}
\label{eq:weyl_state_correction}
\bra[\big]{0} \mathe^{\mathi \phi(f)} \ket[\big]{ \Omega^{(1)} } + \bra[\big]{ \Omega^{(1)} } \mathe^{\mathi \phi(f)} \ket[\big]{0} = \mathe^{- \frac{\mathi}{2} G^+(f,f)} \int\dotsi\int K^{(1)}(x_1,x_2,x_3,x_4) \prod_{j=1}^4 (G^- f)(x_j) \total^4 x_j \eqend{,}
\end{equation}
where $K^{(1)}$ is an integral kernel whose explicit form will be irrelevant for our results.

\section{Relative entropy and the Bekenstein bound}
\label{sec:entropy}

We now have all the ingredients to compute the relative entropy between the interacting Minkowski vacuum state $\ket{\Omega}$ and the interacting coherent state $W(f) \ket{\Omega}$ in the Rindler wedge with $\supp f \subset \mathcal{W}_\mathrm{R}$. By the Bisognano--Wichmann theorem~\cite{bisognanowichmann1975,bisognanowichmann1976}, the modular Hamiltonian of the vacuum is proportional to the generator of boosts, $\ln \Delta_\Omega = 2 \pi \mathcal{M}^{01}$ with the explicit form given in equation~\eqref{eq:minkowski_lorentz_generators}. Since $W(f)$ is a unitary element of $\mathcal{A}$, the relation~\eqref{eq:delta_rel_unitary} holds for the relative modular operator $\Delta_{W(f) \Omega, \Omega}$, and the relative entropy is given by~\eqref{eq:s_rel_unitary}
\begin{splitequation}
\label{eq:srel_omega_wfomega}
S_\mathrm{rel}(\Omega \Vert W(f) \Omega) &= - \bra{\Omega} W(f) \ln \Delta_\Omega [ W(f) ]^\dagger \ket{\Omega} = - 2 \pi \bra{\Omega} W(f) \mathcal{M}^{01} W(-f) \ket{\Omega} \\
&= \mathi \frac{\partial}{\partial s} \bra{\Omega} W(f) \mathe^{2 \pi \mathi s \mathcal{M}^{01}} W(-f) \ket{\Omega} \rvert_{s = 0} \\
&= \mathi \frac{\partial}{\partial s} \bra{\Omega} W(f) \mathe^{2 \pi \mathi s \mathcal{M}^{01}} W(-f) \mathe^{- 2 \pi \mathi s \mathcal{M}^{01}} \ket{\Omega} \rvert_{s = 0} \eqend{,}
\end{splitequation}
where we used that $\mathcal{T}_{\mu\nu}$ and thus $\mathcal{M}^{\mu\nu}$ annihilate the vacuum.

We thus need to compute the expectation of an interacting Weyl operator and a transformed one to first order in the generator. Using the transformation~\eqref{eq:qxi_weyl} of the interacting Weyl operators and the boost Killing vector $\xi^\mu = \eta^{\mu 1} x^0 - \eta^{\mu 0} x^1$ corresponding to $\mathcal{M}^{01}$, we obtain
\begin{equation}
\mathe^{2 \pi \mathi s \mathcal{M}^{01}} W(-f) \mathe^{- 2 \pi \mathi s \mathcal{M}^{01}} = W(-f_s)
\end{equation}
with~\eqref{eq:qxi_phi}
\begin{equation}
f_s(x) = f(x) + 2 \pi s \xi^\mu \partial_\mu f(x) + \bigo{s^2} = f(x) + 2 \pi s L f(x) + \bigo{s^2} \eqend{,}
\end{equation}
where we defined
\begin{equation}
\label{eq:l_def}
L f(x) = x^0 \partial_1 f(x) + x^1 \partial_0 f(x) \eqend{.}
\end{equation}
Using the result~\eqref{eq:weyl_product} for the product of interacting Weyl operators, the expansion~\eqref{eq:interacting_omega_expansion} of the interacting Minkowski vacuum state, and the results~\eqref{eq:weyl_state_correction} for the matrix element and~\eqref{eq:weyl_normalordered_expectation} for the expectation value of interacting Weyl operators, a long but straightforward computation results in
\begin{splitequation}
&\bra{\Omega} W(f) W(-f_s) \ket{\Omega} = \mathe^{- \frac{\mathi}{2} G^+(f,f) + \mathi G^+(f,f_s) - \frac{\mathi}{2} G^+(f_s,f_s)} \\
&\quad\times \biggl[ 1 - \frac{\mathi \lambda}{24} \int ( G^\mathrm{adv}(f-f_s) )(z) \left[ ( G^+ (f-f_s) )(z) + ( G^- (f-f_s) )(z) \right] \\
&\qquad\qquad\quad\times \left[ [ ( G^+ (f-f_s) )(z) ]^2 + [ ( G^- (f-f_s) )(z) ]^2 \right] \total^4 z \\
&\qquad\quad+ \frac{\mathi \lambda}{24} \int \left[ ( G^\mathrm{adv} f )(z) ( \Delta f_s )(z) - ( G^\mathrm{adv} f_s )(z) ( \Delta f )(z) \right] \Bigl[ 6 [ ( G^- (f-f_s) )(z) ]^2 \\
&\qquad\qquad\quad+ 4 \left[ ( \Delta f )(z) - 2 ( \Delta f_s )(z) \right] ( G^- (f-f_s) )(z) + [ ( \Delta f )(z) ]^2 \\
&\qquad\qquad\quad- 3 ( \Delta (f-f_s) )(z) ( \Delta f_s )(z) \Bigr] \total^4 z \\
&\qquad\quad+ \lambda \int\dotsi\int K^{(1)}(x_1,x_2,x_3,x_4) \prod_{j=1}^4 (G^- (f-f_s))(x_j) \total^4 x_j + \bigo{\lambda^2} \biggr] \eqend{,}
\end{splitequation}
where we defined in analogy to~\eqref{eq:delta_fg_def}
\begin{equation}
G^+(f,g) = \iint f(x) G^+(x,y) g(y) \total^4 x \total^4 y = \int f(x) ( G^+ g )(x) \total^4 x \eqend{.}
\end{equation}

Inserting this result in~\eqref{eq:srel_omega_wfomega}, it follows that
\begin{splitequation}
S_\mathrm{rel}(\Omega \Vert W(f) \Omega) &= \frac{\partial}{\partial s} \left[ - G^+(f,f_s) + \frac{1}{2} G^+(f_s,f_s) \right]_{s = 0} \\
&\quad- \frac{\lambda}{24} \int \frac{\partial}{\partial s} \left[ ( G^\mathrm{adv} f )(z) ( \Delta f_s )(z) - ( G^\mathrm{adv} f_s )(z) ( \Delta f )(z) \right]_{s = 0} [ ( \Delta f )(z) ]^2 \total^4 z \\
&\quad+ \bigo{\lambda^2} \eqend{,}
\end{splitequation}
and in particular we see that the correction to the free Minkowski vacuum does not contribute at first order. For the individual terms, we compute
\begin{splitequation}
\label{eq:s_der_combination_1}
\frac{\partial}{\partial s} \left[ - G^+(f,f_s) + \frac{1}{2} G^+(f_s,f_s) \right]_{s = 0} &= - \frac{1}{2} \iint f(x) \Delta(x,y) \frac{\partial}{\partial s} f_s(y) \rvert_{s = 0} \total^4 x \total^4 y \\
&= - \pi \iint f(x) \Delta(x,y) (L f)(y) \total^4 x \total^4 y \\
&= \pi \int L f(x) ( \Delta f )(x) \total^4 x \eqend{,}
\end{splitequation}
as well as
\begin{splitequation}
\label{eq:s_der_combination_2}
&\frac{\partial}{\partial s} \left[ ( G^\mathrm{adv} f )(z) ( \Delta f_s )(z) - ( G^\mathrm{adv} f_s )(z) ( \Delta f )(z) \right]_{s = 0} \\
&\quad= 2 \pi ( G^\mathrm{adv} f )(z) \int \Delta(z,x) L f(x) \total^4 x - 2 \pi \int G^\mathrm{adv}(z,x) L f(x) \total^4 x ( \Delta f )(z) \\
&\quad= 2 \pi \left[ ( G^\mathrm{adv} f )(z) L ( \Delta f )(z) - L ( G^\mathrm{adv} f )(z) ( \Delta f )(z) \right] \\
&\quad= 2 \pi \left[ ( G^\mathrm{ret} f )(z) L ( \Delta f )(z) - ( \Delta f )(z) L ( G^\mathrm{ret} f )(z) \right] \eqend{,}
\end{splitequation}
where we used that~\eqref{eq:app_l_commute_delta}, \eqref{eq:app_l_commute_gretadv}
\begin{equation}
( \Delta L f )(x) = L ( \Delta f )(x) \eqend{,} \quad ( G^\mathrm{adv} L f )(x) = L ( G^\mathrm{adv} f )(x)
\end{equation}
for compactly supported $f$, as well as the definition $\Delta(x,y) = G^\mathrm{ret}(x,y) - G^\mathrm{adv}(x,y)$~\eqref{eq:delta_def} of the commutator function. We therefore obtain
\begin{splitequation}
\label{eq:srel_expression1}
S_\mathrm{rel}(\Omega \Vert W(f) \Omega) &= \pi \int L f(x) ( \Delta f )(x) \total^4 x \\
&\quad- \frac{\pi \lambda}{12} \int \left[ ( G^\mathrm{ret} f )(z) L ( \Delta f )(z) - ( \Delta f )(z) L ( G^\mathrm{ret} f )(z) \right] [ ( \Delta f )(z) ]^2 \total^4 z \\
&\quad+ \bigo{\lambda^2} \\
&= \pi \int L f(x) ( \Delta f )(x) \total^4 x - \frac{\pi \lambda}{3} \int ( G^\mathrm{ret} f )(x) [ ( \Delta f )(x) ]^2 L ( \Delta f )(x) \total^4 x \\
&\quad+ \frac{\pi \lambda}{12} \int L \left[ ( G^\mathrm{ret} f )(x) [ ( \Delta f )(x) ]^3 \right] \total^4 x + \bigo{\lambda^2} \eqend{.}
\end{splitequation}

We now would like to bring this into a form where positivity is manifest, analogously to the case of free fields~\cite{casinigrillopontello2019,longo2019,ciollilongoruzzi2020,froebmuchpapadopoulos2023}. For this, we first note that the last term in~\eqref{eq:srel_expression1} simplifies to
\begin{splitequation}
\label{eq:srel_expression1_term3}
\int L \left[ ( G^\mathrm{ret} f)(x) [ ( \Delta f )(x) ]^3 \right] \total^4 x &= \int x^0 \partial_1 \left[ ( G^\mathrm{ret} f )(x) [ ( \Delta f )(x) ]^3 \right] \total^4 x \\
&\qquad+ \int x^1 \partial_0 \left[ ( G^\mathrm{ret} f )(x) [ ( \Delta f )(x) ]^3 \right] \total^4 x \\
&= \lim_{t \to \infty} \int_{x^0 = t} x^1 ( G^\mathrm{ret} f )(x) [ ( \Delta f )(x) ]^3 \total^3 \vec{x} \\
&= \lim_{t \to \infty} \int_{x^0 = t} x^1 [ ( \Delta f )(x) ]^4 \total^3 \vec{x} \eqend{,}
\end{splitequation}
because $G^\mathrm{ret/adv} f$ and thus also $\Delta f$ are spatially compact for compactly supported $f$, i.e., for any fixed time $x^0 = t$ the spatial support is compact, and where the last equality follows because $( G^\mathrm{adv} f )(x) = ( G^\mathrm{ret} f )(x) - ( \Delta f )(x)$ vanishes for all $x \not\in J^-( \supp f )$ for compactly supported $f$, in particular in the limit $t \to \infty$. On the other hand, for the first term in~\eqref{eq:srel_expression1} we employ the classical Green's identity~\eqref{eq:app_green_identity}
\begin{equation}
\label{eq:green_identity}
\int f(x) ( \Delta g )(x) \total^4 x = \int_{x^0 = t} \Bigl[ ( \Delta f )(x) \partial_0 ( \Delta g )(x) - ( \Delta g )(x) \partial_0 ( \Delta f )(x) \Bigr] \total^3 \vec{x} \eqend{,}
\end{equation}
valid for compactly supported $f$ and arbitrary $t$. Choosing $t = 0$, this yields
\begin{splitequation}
\label{eq:lf_deltaf_reduction}
\int L f(x) ( \Delta f )(x) \total^4 x &= \int_{x^0 = 0} \Bigl[ ( \Delta L f )(x) \partial_0 ( \Delta f )(x) - ( \Delta f )(x) \partial_0 ( \Delta L f )(x) \Bigr] \total^3 \vec{x} \\
&= \int_{x^0 = 0} \Bigl[ x^1 \partial_0 ( \Delta f )(x) \partial_0 ( \Delta f )(x) - ( \Delta f )(x) \partial_1 ( \Delta f )(x) \\
&\qquad\qquad- ( \Delta f )(x) x^1 \partial_0^2 ( \Delta f )(x) \Bigr] \total^3 \vec{x} \\
&= \int_{x^0 = 0} \Bigl[ x^1 \partial_0 ( \Delta f )(x) \partial_0 ( \Delta f )(x) - ( \Delta f )(x) \partial_1 ( \Delta f )(x) \\
&\qquad\qquad- x^1 ( \Delta f )(x) \partial^k \partial_k ( \Delta f )(x) + x^1 m^2 ( \Delta f )(x) ( \Delta f )(x) \Bigr] \total^3 \vec{x} \\
&= \int_{x^0 = 0} x^1 \Bigl[ \partial_0 ( \Delta f )(x) \partial_0 ( \Delta f )(x) + \partial^k ( \Delta f )(x) \partial_k ( \Delta f )(x) \\
&\qquad\qquad+ m^2 ( \Delta f )(x) ( \Delta f )(x) \Bigr] \total^3 \vec{x} \eqend{,}
\end{splitequation}
using that~\eqref{eq:app_l_commute_delta} $( \Delta L f )(x) = L ( \Delta f )(x)$, computing the derivatives explicitly, and finally using that the commutator function $\Delta$ is a bisolution of the Klein--Gordon equation and integrating spatial derivatives by parts (where no boundary terms arise because $\Delta f$ is space-compact).

For the remaining second term in~\eqref{eq:srel_expression1}, we cannot use the identity~\eqref{eq:green_identity} since $G^\mathrm{ret} f$ and $\Delta f$ are only space-compact for compactly supported $f$, but have to use the extended identity~\eqref{eq:app_green_spacecompact}
\begin{splitequation}
\label{eq:green_spacecompact}
\int f(x) ( \Delta g )(x) \total^4 x &= \int_{x^0 = 0} \left[ ( \Delta f )(x) \partial_0 ( \Delta g )(x) - \partial_0 ( \Delta f )(x) ( \Delta g )(x) \right] \total^3 \vec{x} \\
&\quad+ \lim_{t \to \infty} \int_{x^0 = t} \left[ ( G^\mathrm{adv} f )(x) \partial_0 ( \Delta g )(x) - \partial_0 ( G^\mathrm{adv} f )(x) ( \Delta g )(x) \right] \total^3 \vec{x} \\
&\quad- \lim_{t \to - \infty} \int_{x^0 = t} \left[ ( G^\mathrm{ret} f )(x) \partial_0 ( \Delta g )(x) - \partial_0 ( G^\mathrm{ret} f )(x) ( \Delta g )(x) \right] \total^3 \vec{x} \eqend{,}
\end{splitequation}
which we derive in appendix~\ref{app:green}. Choosing $h(x) = ( G^\mathrm{ret} f )(x) [ ( \Delta f )(x) ]^2$ instead of $f(x)$ and $g(x) = L f(x)$, this gives
\begin{splitequation}
&\int ( G^\mathrm{ret} f )(x) [ ( \Delta f )(x) ]^2 ( \Delta L f )(x) \total^4 x \\
&\quad= \int_{x^0 = 0} \left[ ( \Delta h )(x) \partial_0 ( \Delta L f )(x) - \partial_0 ( \Delta h )(x) ( \Delta L f )(x) \right] \total^3 \vec{x} \\
&\qquad+ \lim_{t \to \infty} \int_{x^0 = t} \left[ ( G^\mathrm{adv} h )(x) \partial_0 ( \Delta L f )(x) - \partial_0 ( G^\mathrm{adv} h )(x) ( \Delta L f )(x) \right] \total^3 \vec{x} \\
&\qquad- \lim_{t \to - \infty} \int_{x^0 = t} \left[ ( G^\mathrm{ret} h )(x) \partial_0 ( \Delta L f )(x) - \partial_0 ( G^\mathrm{ret} h )(x) ( \Delta L f )(x) \right] \total^3 \vec{x} \eqend{,}
\end{splitequation}
and manipulations analogous to the ones performed in~\eqref{eq:lf_deltaf_reduction} show that the first integral reduces to
\begin{splitequation}
&\int_{x^0 = 0} \left[ ( \Delta h )(x) \partial_0 ( \Delta L f )(x) - \partial_0 ( \Delta h )(x) ( \Delta L f )(x) \right] \total^3 \vec{x} \\
&\quad= - \int_{x^0 = 0} x^1 \left[ \partial_0 ( \Delta h )(x) \partial_0 ( \Delta f )(x) + \partial^k ( \Delta h )(x) \partial_k ( \Delta f )(x) + m^2 ( \Delta h )(x) ( \Delta f )(x) \right] \total^3 \vec{x} \eqend{.}
\end{splitequation}
Moreover, we have
\begin{equation}
( G^\mathrm{adv} h )(x) = ( G^\mathrm{adv} [ ( G^\mathrm{ret} f ) ( \Delta f )^2 ] )(x) = ( G^\mathrm{adv} [ ( \Delta f )^3 ] )(x) + ( G^\mathrm{adv} [ ( G^\mathrm{adv} f ) ( \Delta f )^2 ] )(x)
\end{equation}
and
\begin{equation}
( G^\mathrm{adv} [ ( G^\mathrm{adv} f ) ( \Delta f )^2 ] )(x) = \iint G^\mathrm{adv}(x,y) G^\mathrm{adv}(y,z) f(z) [ ( \Delta f )(y) ]^2 \total^4 x \total^4 y \eqend{,}
\end{equation}
and since $\supp ( G^\mathrm{adv} g ) \subseteq J^-( \supp g )$ and $f$ is compactly supported, it follows that \linebreak $\lim_{x^0 \to \infty} ( G^\mathrm{adv} [ ( G^\mathrm{adv} f ) ( \Delta f )^2 ] )(x) = 0$. Analogously, one shows that
\begin{equation}
\lim_{x^0 \to - \infty} ( G^\mathrm{ret} h )(x) = \lim_{x^0 \to - \infty} \iint G^\mathrm{ret}(x,y) G^\mathrm{ret}(y,z) f(z) [ ( \Delta f )(y) ]^2 \total^4 x \total^4 y = 0 \eqend{,}
\end{equation}
such that
\begin{splitequation}
\label{eq:srel_expression1_term2a}
&\int ( G^\mathrm{ret} f )(x) [ ( \Delta f )(x) ]^2 ( \Delta L f )(x) \total^4 x \\
&\quad= - \int_{x^0 = 0} x^1 \left[ \partial_0 ( \Delta h )(x) \partial_0 ( \Delta f )(x) + \partial^k ( \Delta h )(x) \partial_k ( \Delta f )(x) + m^2 ( \Delta h )(x) ( \Delta f )(x) \right] \total^3 \vec{x} \\
&\qquad+ \lim_{t \to \infty} \int_{x^0 = t} \left[ ( G^\mathrm{adv} [ ( \Delta f )^3 ] )(x) \partial_0 ( \Delta L f )(x) - \partial_0 ( G^\mathrm{adv} [ ( \Delta f )^3 ] )(x) ( \Delta L f )(x) \right] \total^3 \vec{x} \eqend{.}
\end{splitequation}

Using that $\Delta$ is a bisolution and $G^\mathrm{adv}$ is a Green's function for the free Klein--Gordon equation, a short computation then yields
\begin{splitequation}
&\partial_0 \left[ ( G^\mathrm{adv} [ ( \Delta f )^3 ] )(x) \partial_0 ( \Delta L f )(x) - \partial_0 ( G^\mathrm{adv} [ ( \Delta f )^3 ] )(x) ( \Delta L f )(x) \right] \\
&\quad= \partial^k \left[ ( G^\mathrm{adv} [ ( \Delta f )^3 ] )(x) \partial_k L ( \Delta f )(x) - \partial_k ( G^\mathrm{adv} [ ( \Delta f )^3 ] )(x) L ( \Delta f )(x) \right] + \frac{1}{4} L [ ( \Delta f )(x) ]^4 \eqend{.}
\end{splitequation}
Since $\Delta f$ is space-compact, the first term on the right-hand side vanishes after integration over a surface of constant time, and integrating the result between $x^0 = 0$ and $x^0 = t$ we obtain
\begin{splitequation}
&\int_{x^0 = t} \left[ ( G^\mathrm{adv} [ ( \Delta f )^3 ] )(x) \partial_0 ( \Delta L f )(x) - \partial_0 ( G^\mathrm{adv} [ ( \Delta f )^3 ] )(x) ( \Delta L f )(x) \right] \total^3 \vec{x} \\
&\quad= \int_{x^0 = 0} \left[ ( G^\mathrm{adv} [ ( \Delta f )^3 ] )(x) \partial_0 ( \Delta L f )(x) - \partial_0 ( G^\mathrm{adv} [ ( \Delta f )^3 ] )(x) ( \Delta L f )(x) \right] \total^3 \vec{x} \\
&\qquad+ \frac{1}{4} \int_{x^0 \in [0,t]} L [ ( \Delta f )(x) ]^4 \total^4 x \eqend{,}
\end{splitequation}
such that~\eqref{eq:srel_expression1_term2a} reduces to
\begin{splitequation}
&\int ( G^\mathrm{ret} f )(x) [ ( \Delta f )(x) ]^2 ( \Delta L f )(x) \total^4 x \\
&\quad= - \int_{x^0 = 0} x^1 \left[ \partial_0 ( \Delta h )(x) \partial_0 ( \Delta f )(x) + \partial^k ( \Delta h )(x) \partial_k ( \Delta f )(x) + m^2 ( \Delta h )(x) ( \Delta f )(x) \right] \total^3 \vec{x} \\
&\qquad+ \int_{x^0 = 0} \left[ ( G^\mathrm{adv} [ ( \Delta f )^3 ] )(x) \partial_0 ( \Delta L f )(x) - \partial_0 ( G^\mathrm{adv} [ ( \Delta f )^3 ] )(x) ( \Delta L f )(x) \right] \total^3 \vec{x} \\
&\qquad+ \frac{1}{4} \int_{x^0 \in [0,\infty)} L [ ( \Delta f )(x) ]^4 \total^4 x \eqend{.}
\end{splitequation}
Performing now manipulations analogous to the ones in~\eqref{eq:lf_deltaf_reduction}, and using the definition $\Delta(x,y) = G^\mathrm{ret}(x,y) - G^\mathrm{adv}(x,y)$~\eqref{eq:delta_def} of the commutator function to combine terms
\begin{equation}
( \Delta h )(x) + ( G^\mathrm{adv} [ ( \Delta f )^3 ] )(x) = ( G^\mathrm{ret} [ ( G^\mathrm{ret} f ) ( \Delta f )^2 ] )(x) - ( G^\mathrm{adv} [ ( G^\mathrm{adv} f ) ( \Delta f )^2 ] )(x) \equiv g(x) \eqend{,}
\end{equation}
this reduces to
\begin{splitequation}
\label{eq:srel_expression1_term2b}
&\int ( G^\mathrm{ret} f )(x) [ ( \Delta f )(x) ]^2 ( \Delta L f )(x) \total^4 x \\
&\quad= - \int_{x^0 = 0} x^1 \left[ \partial_0 g(x) \partial_0 ( \Delta f )(x) + \partial^k g(x) \partial_k ( \Delta f )(x) + m^2 g(x) ( \Delta f )(x) \right] \total^3 \vec{x} \\
&\qquad- \frac{1}{4} \int_{x^0 \in [0,\infty)} L [ ( \Delta f )(x) ]^4 \total^4 x \eqend{.}
\end{splitequation}
Inserting the results~\eqref{eq:srel_expression1_term3}, \eqref{eq:lf_deltaf_reduction} and~\eqref{eq:srel_expression1_term2b}, it follows that the relative entropy~\eqref{eq:srel_expression1} can be written as
\begin{splitequation}
\label{eq:srel_expression2}
S_\mathrm{rel}(\Omega \Vert W(f) \Omega) &= \pi \int_{x^0 = 0} x^1 \Bigl[ \partial_0 ( \Delta f )(x) \partial_0 ( \Delta f )(x) + \partial^k ( \Delta f )(x) \partial_k ( \Delta f )(x) \\
&\hspace{6em}+ m^2 ( \Delta f )(x) ( \Delta f )(x) \Bigr] \total^3 \vec{x} \\
&\quad+ \frac{\pi \lambda}{3} \int_{x^0 = 0} x^1 \left[ \partial_0 g(x) \partial_0 ( \Delta f )(x) + \partial^k g(x) \partial_k ( \Delta f )(x) + m^2 g(x) ( \Delta f )(x) \right] \total^3 \vec{x} \\
&\quad- \frac{\pi \lambda}{12} \int_{x^0 \in [0,\infty)} L [ ( \Delta f )(x) ]^4 \total^4 x + \frac{\pi \lambda}{12} \lim_{t \to \infty} \int_{x^0 = t} x^1 [ ( \Delta f )(x) ]^4 \total^3 \vec{x} + \bigo{\lambda^2} \eqend{.}
\end{splitequation}

Using that $\Delta f$ is spatially compact, we obtain that the last two terms combine to
\begin{equation}
- \int_{x^0 \in [0,\infty)} L [ ( \Delta f )(x) ]^4 \total^4 x + \lim_{t \to \infty} \int_{x^0 = t} x^1 [ ( \Delta f )(x) ]^4 \total^3 \vec{x} = \int_{x^0 = 0} x^1 [ ( \Delta f )(x) ]^4 \total^4 x \eqend{,}
\end{equation}
and it follows that the relative entropy~\eqref{eq:srel_expression2} reads
\begin{equation}
\label{eq:srel_stress}
S_\mathrm{rel}(\Omega \Vert W(f) \Omega) = 2 \pi \int_{x^0 = 0} x^1 T_{00}\left( \varphi^\mathrm{int}_f \right) \total^3 \vec{x} + \bigo{\lambda^2} \eqend{,}
\end{equation}
where
\begin{equation}
\label{eq:energy_density}
T_{00}(\varphi) = \frac{1}{2} \left( \partial_0 \varphi \right)^2 + \frac{1}{2} \partial^k \varphi \partial_k \varphi + \frac{1}{2} m^2 \varphi^2 + \frac{\lambda}{4!} \varphi^4
\end{equation}
is the classical energy density of the interacting theory, and
\begin{splitequation}
\varphi^\mathrm{int}_f(x) &= ( \Delta f )(x) + \frac{\lambda}{6} g(x) + \bigo{\lambda^2} \\
&= ( \Delta f )(x) + \frac{\lambda}{6} ( G^\mathrm{ret} [ ( G^\mathrm{ret} f ) ( \Delta f )^2 ] )(x) - \frac{\lambda}{6} ( G^\mathrm{adv} [ ( G^\mathrm{adv} f ) ( \Delta f )^2 ] )(x) + \bigo{\lambda^2}
\end{splitequation}
is the classical causal interacting solution~\eqref{eq:phiclass_int_def} of the Klein--Gordon equation, which satisfies the support property $\supp \varphi^\mathrm{int}_f \subseteq J^+(\supp f) \cup J^-(\supp f)$. In particular, since we chose $f$ such that $\supp f \subset \mathcal{W}_\mathrm{R}$, we have $\supp \varphi^\mathrm{int}_f \rvert_{x^0 = 0} \subset \mathcal{W}_\mathrm{R} \rvert_{x^0 = 0} = \{ x \in \mathbb{R}^{1,d-1} \colon x^0 = 0, x^1 \geq 0 \}$, and the integral in the result~\eqref{eq:srel_stress} for the relative entropy only ranges over $x^1 \geq 0$, such that the relative entropy is manifestly positive: $S_\mathrm{rel}(\Omega \Vert W(f) \Omega) \geq 0$.

\subsection{Bekenstein bound}
\label{sec:bekenstein}

According to Bekenstein~\cite{bekenstein1981}, the entropy $S$ of any system of finite extent $R$ is bounded by $2 \pi R E$, where $E$ is its energy. While this bound was derived using a Gedankenexperiment involving black holes, it applies to nongravitational systems in flat space; a covariant generalization was given by Bousso~\cite{bousso1999}. However, since the local algebras of quantum field theory are of type III~\cite{fredenhagen1985,buchholzdantonifredenhagen1987}, the entanglement entropy between the system and its environment, given as the von Neumann entropy of the reduced density matrix of the system, is UV-divergent~\cite{bombellikoulleesorkin1986}, and only relative entropy remains finite. Therefore, a rigorous formulation of the Bekenstein bound can be stated as follows:\footnote{A different formulation of the Bekenstein bound was given by Casini~\cite{casini2008}, which reduces to the positivity of relative entropy and thus holds always~\cite{longoxu2018b,kudlerflametal2025}.} the relative entropy between the vacuum and an excited state is bounded by $2 \pi R E$, when the excitation is confined to a region of diameter $2 R$. Using nontrivial estimates for the relative modular Hamiltonian for a general excited state~\cite{buchholzdantonilongo1990,buchholzdantonilongo2007}, this bound has been proven by Longo~\cite{longo2025}, and analogous but weaker bounds have been proven by Hollands and Longo for excitations that are only approximately localized in the given region~\cite{hollandslongo2025a}.

Remarkably, in the case that we consider in this paper the bound is much easier to obtain. Namely, by a computation fully analogous to the one for relative entropy in the last subsection, we obtain
\begin{equation}
E\Bigl( W(f) \Omega \Bigr) = \bra{\Omega} W(-f) \mathcal{P}^0 W(f) \ket{\Omega} = \int_{x^0 = 0} T_{00}\left( \varphi^\mathrm{int}_f \right) \total^3 \vec{x} + \bigo{\lambda^2} \eqend{,}
\end{equation}
where $\mathcal{P}^0$ is the Hamiltonian~\eqref{eq:minkowski_lorentz_generators}, $T_{00}$ is the classical energy density~\eqref{eq:energy_density} and $\varphi^\mathrm{int}_f$ is the classical causal interacting solution~\eqref{eq:phiclass_int_def} of the Klein--Gordon equation. Since the computation is identical to the one for relative entropy, with the only change being the replacement of $L$~\eqref{eq:l_def} by $\partial_0$, we refrain from giving any details. Assuming now that $\supp f \rvert_{x^0 = 0} \subseteq \{ x \colon x^0 = 0, x^1 \in [0,R] \}$, by the support property $\supp \varphi^\mathrm{int}_f \subseteq J^+(\supp f) \cup J^-(\supp f)$ of the classical causal interaction solution we also have $\supp \varphi^\mathrm{int}_f \rvert_{x^0 = 0} \subseteq \{ x \colon x^0 = 0, x^1 \in [0,R] \}$. Since the classical energy density $T_{00}$ is positive, we obtain from the result~\eqref{eq:srel_stress}
\begin{splitequation}
\label{eq:srel_bekenstein}
S_\mathrm{rel}(\Omega \Vert W(f) \Omega) &= 2 \pi \int_{x^0 = 0, x^1 \in [0,R]} x^1 T_{00}\left( \varphi^\mathrm{int}_f \right) \total^3 \vec{x} + \bigo{\lambda^2} \\
&\leq 2 \pi R \int_{x^0 = 0, x^1 \in [0,R]} T_{00}\left( \varphi^\mathrm{int}_f \right) \total^3 \vec{x} + \bigo{\lambda^2} = 2 \pi R E\Bigl( W(f) \Omega \Bigr) \eqend{,}
\end{splitequation}
which is exactly the Bekenstein bound for relative entropy.

\section{Conclusion and outlook}
\label{sec:conclusion}

We have considered the relative entropy $S_\mathrm{rel}(\Omega \Vert W(f) \Omega)$ between the Minkowski vacuum state $\ket{\Omega}$ and the interacting coherent state $W(f) \ket{\Omega}$ in the Rindler wedge with $\supp f \subset \mathcal{W}_\mathrm{R}$. For this, we have constructed the interacting Weyl algebra generated by the interacting Weyl operators $W(f) = \mathe^{\mathi \Phi(f)}$, obtained in turn from the interacting field operators $\Phi(f)$. We proved that its restriction to the right Rindler wedge, choosing $\supp f \subset \mathcal{W}_\mathrm{R}$, respects Haag duality such that the commutant of the right-wedge algebra is the left-wedge algebra, which is generated by interacting Weyl operators $W(f)$ with $\supp f \subset \mathcal{W}_\mathrm{L}$. Therefore, the Bisognano--Wichmann theorem is applicable, and the relative entropy $S_\mathrm{rel}(\Omega \Vert W(f) \Omega)$ can be computed using the modular Hamiltonian $\ln \Delta_\omega = 2 \pi \mathcal{M}^{01}$ with the interacting boost generator $\mathcal{M}^{01}$ defined using the interacting stress tensor $\mathcal{T}_{\mu\nu}$. We have then shown that to first order in the coupling $\lambda$, the relative entropy is given by~\eqref{eq:srel_stress}
\begin{equation}
S_\mathrm{rel}(\Omega \Vert W(f) \Omega) = 2 \pi \int_{x^0 = 0} x^1 T_{00}\left( \varphi^\mathrm{int}_f \right) \total^3 \vec{x} + \bigo{\lambda^2} \eqend{,}
\end{equation}
and is equal to the \emph{classical} interacting Noether charge $- M^{01} = M^{10}$~\eqref{eq:classical_noether_m} (up to a factor of $2 \pi$) associated to the boosts, evaluated on the classical interacting solution $\varphi^\mathrm{int}_f$~\eqref{eq:phiclass_int_def}. This generalizes the corresponding result in the free theory~\cite{casinigrillopontello2019,froebmuchpapadopoulos2023}, and maybe was to be expected, but as we have seen, the computation is quite nontrivial. Moreover, we have shown that the classical solution on which the charge is evaluated is causally evolving, i.e., with $\supp \varphi^\mathrm{int}_f \subseteq J^+(\supp f) \cup J^-(\supp f)$, which was not clear a priori. In fact, since this solution is not uniquely determined by this support property (unlike the retarded or advanced solutions), it is necessary to determine it by the explicit computation. We can thus conjecture in good conscience that
\begin{equation}
S_\mathrm{rel}(\Omega \Vert W(f) \Omega) = 2 \pi \int_{x^0 = 0} x^1 T_{00}\left( \varphi^\mathrm{int}_f \right) \total^3 \vec{x}
\end{equation}
holds to all orders in $\lambda$, where $\varphi^\mathrm{int}_f$ is a causal interaction solution of the interacting Klein--Gordon equation. However, without further computation, we cannot determine $\varphi^\mathrm{int}_f$ exactly, since it is not unique. Because the classical energy density $T_{00}$ is positive, for a function $f$ supported in a strip of width $R$, namely $\supp f \rvert_{x^0 = 0} \subseteq \{ x \colon x^0 = 0, x^1 \in [0,R] \}$, we can simply estimate $x^1$ by $R$ and obtain the Bekenstein bound~\eqref{eq:srel_bekenstein}
\begin{equation}
S_\mathrm{rel}(\Omega \Vert W(f) \Omega) \leq 2 \pi R \int_{x^0 = 0, x^1 \in [0,R]} T_{00}\left( \varphi^\mathrm{int}_f \right) \total^3 \vec{x} + \bigo{\lambda^2} = 2 \pi R E\Bigl( W(f) \Omega \Bigr) \eqend{,}
\end{equation}
which we again conjecture to hold to all orders in $\lambda$.

Apart from explicitly computing the relative entropy and verifying the Bekenstein bound, our results can now be used for further studies in perturbatively interacting $\lambda \phi^4$ theory. A relatively straightforward extension would be the computation of Petz--Rényi entropies~\cite{petz1985,lashkari2019,froebsangaletti2025}, or $f$-divergences that can be defined for an arbitrary convex function $f$ in terms of the relative modular Hamiltonian~\cite{petz1985,hiai2018,hiai2019}. Other directions include the determination of the capacity of entanglement, the analog of heat capacity in thermodynamics, which is the variance of the modular Hamiltonian. In statistical physics, this quantity can be used to distinguish states with and without topologically protected gapless entanglement spectrum~\cite{yaoqi2010}, and in quantum field theory it has been studied only quite recently (see, for example, \cite{nakagawafurukawa2017,banerjeeerdmengersarkar2018,deboerjarvelakeskivakkuri2019,shrimalibhowmickpandeypati2022,ariasdigiuliokeskivakkuritonni2023,andrzejewski2023,mohammadimozaffar2024,aalsmabak2025,bubsivaramakrishnan2025}). In particular, in some cases it is possible to relate capacity of entanglement to the entanglement entropy itself~\cite{banksdraper2024}, and possible to obtain bounds reminiscent of the Bekenstein bound. Moreover, considering higher moments of the modular Hamiltonian, it is possible to construct infinite sequences of so-called entanglement monotones~\cite{ariasdeboerdigiuliokeskivakkuritonni2023}, functions that like the relative entropy decrease under the application of quantum channels. Also extensions to other theories should be relatively straightforward, for example, a Yukawa model also involving fermions; the relative entropy for free fermions is also known in various situations (see, for example, \cite{longoxu2018a,friesreyes2019,xu2023,galandamuchverch2023,finstermuch2025} and references therein).

Another extension concerns regions different from the Rindler wedge in Minkowski spacetime, for example the exterior of a Schwarzschild black hole where the modular Hamiltonian of the Hartle--Hawking state is also known~\cite{sewell1982,kay1985a,kay1985b,kaywald1991}. In fact, the modular Hamiltonian is known for the generalized Unruh state on any bifurcate Killing horizon~\cite{summersverch1996} (see also~\cite{kurpiczpinamontiverch2021}), where it acts by dilations along the horizon. One can thus use modular theory to compute the relative entropy between this state and a coherent excitation thereof, and furthermore study the classical backreaction that is exerted by the classical solution corresponding to the expectation value of the field operator in the coherent state on the background geometry, leading to rigorous entropy-area laws that relate relative entropy with a change in horizon area~\cite{hollandsishibashi2019,dangelo2020,dangelo2021} (see also~\cite{dangeloetal2023} for a similar computation in de Sitter spacetime).\footnote{These results were derived under some assumptions of compact support. For an extension without these assumptions, which is required to study soft modes that do not decay asymptotically, see~\cite{danielsonsatishchandran2025}.} Relative entropy has been further used in these settings to derive the semiclassical Einstein equations~\cite{doraumuch2026} and the quantum null energy condition~\cite{ceyhanfaulkner2020,morinellitanimotowegener2022,hollandslongo2025b}\footnote{See also the recent~\cite{flissrolph2025,bendayansrivastava2026} for further extensions.}, as well as $c$, $f$, and $a$ theorems~\cite{casinisalazarlandeatorroba2023,abatetorroba2024}. While many of these results follow from general properties of the relative entropy and are independent of the concrete form of the modular Hamiltonian, it would be interesting to see, e.g. if bounds become tighter as one includes interactions.

Finally, we note that so far we have been considering only static situations, in the sense that both states are fixed. It is, of course, interesting to see how relative entropy changes when one state is evolved in time, and in particular study the rate of entropy production. A general formula for entropy production has been given for quantum statistical mechanics and related to relative entropy in~\cite{jaksicpillet2001,jaksicpillet2002,jaksicpillet2003} under the assumption that one state is invariant under the evolution, and an analogous formula was given for perturbatively interacting field theories in~\cite{dragofaldinopinamonti2018b}. A vanishing entropy production characterizes steady states, which, however, do not need to be equilibrium states; nevertheless, in many cases perturbed states return to an equilibrium state at late times~\cite{doyonlucasschalmbhaseen2015,dragofaldinopinamonti2018a}. An interesting application of this formalism to gravitational algebras was given recently~\cite{cirafici2024}, which might make it possible to study black hole evaporation in a rigorous way.

\begin{acknowledgments}
This work was supported and funded by \emph{Kuwait University}, Research Project No.~SM05/24.
\end{acknowledgments}

\subsection*{Conflict of interest statement}

On behalf of all authors, the corresponding author states that there is no conflict of interest.

\subsection*{Data availability statement}

This is a purely mathematical work and no data was created or analyzed in this study.

\appendix

\section{Renormalization of a product of propagators}
\label{app:renormalization}

We need to define renormalized versions of a product of two Feynman propagators and their derivatives, fulfilling the condition~\eqref{eq:stress_conserved_condition}
\begin{equation}
\label{eq:app_condition}
2 \partial^\mu \left[ \partial_\mu G^\mathrm{F}(x,y) \partial_\nu G^\mathrm{F}(x,y) \right]_\mathrm{ren} - \partial_\nu \left[ \partial_\mu G^\mathrm{F}(x,y) \partial^\mu G^\mathrm{F}(x,y) \right]_\mathrm{ren} - m^2 \partial_\nu \left[ G^\mathrm{F}(x,y) \right]^2_\mathrm{ren} = 0
\end{equation}
to obtain a conserved renormalized stress tensor. This is most easily done in Fourier space, where we have
\begin{equation}
G^\mathrm{F}(x,y) = - \int \frac{\mathe^{\mathi p (x-y)}}{p^2 + m^2 - \mathi \epsilon} \frac{\total^d p}{(2 \pi)^d} \eqend{,}
\end{equation}
and thus formally
\begin{splitequation}
\left[ G^\mathrm{F}(x,y) \right]^2 &= \iint \frac{\mathe^{\mathi (p+q) (x-y)}}{( p^2 + m^2 - \mathi \epsilon ) ( q^2 + m^2 - \mathi \epsilon )} \frac{\total^d p}{(2 \pi)^d} \frac{\total^d q}{(2 \pi)^d} \\
&= \int \mathe^{\mathi p (x-y)} \left[ \int \frac{1}{[ (p-q)^2 + m^2 - \mathi \epsilon ] ( q^2 + m^2 - \mathi \epsilon )} \frac{\total^d q}{(2 \pi)^d} \right] \frac{\total^d p}{(2 \pi)^d} \eqend{,}
\end{splitequation}
where the inner integral over $q$ is logarithmically divergent for $d = 4$. We use the Feynman parameter formula
\begin{equation}
\frac{1}{A B} = \int_0^1 \frac{1}{\left[ \xi A + (1-\xi) B \right]^2} \total \xi
\end{equation}
and obtain
\begin{splitequation}
&\int \frac{1}{[ (p-q)^2 + m^2 - \mathi \epsilon ] ( q^2 + m^2 - \mathi \epsilon )} \frac{\total^d q}{(2 \pi)^d} \\
&\quad= \int_0^1 \int \frac{1}{\left[ \xi (p-q)^2 + (1-\xi) q^2 + m^2 - \mathi \epsilon \right]^2} \frac{\total^d q}{(2 \pi)^d} \total \xi \\
&\quad= \int_0^1 \int \frac{1}{\left[ q^2 + \xi (1-\xi) p^2 + m^2 - \mathi \epsilon \right]^2} \frac{\total^d q}{(2 \pi)^d} \total \xi \\
&\quad= \int_0^1 \int \left[ \frac{1}{\left[ q^2 + \xi (1-\xi) p^2 + m^2 - \mathi \epsilon \right]^2} - \frac{1}{\left( q^2 + m^2 - \mathi \epsilon \right)^2} \right] \frac{\total^d q}{(2 \pi)^d} \total \xi \\
&\qquad+ \int \frac{1}{\left( q^2 + m^2 - \mathi \epsilon \right)^2} \frac{\total^d q}{(2 \pi)^d} \eqend{.}
\end{splitequation}
Replacing the last (divergent) term by a finite constant $c$, we define
\begin{splitequation}
\left[ G^\mathrm{F}(x,y) \right]^2_\mathrm{ren} &= \lim_{d \to 4} \int \mathe^{\mathi p (x-y)} \int_0^1 \int \Biggl[ \frac{1}{\left[ q^2 + \xi (1-\xi) p^2 + m^2 - \mathi \epsilon \right]^2} \\
&\hspace{10em}- \frac{1}{\left( q^2 + m^2 - \mathi \epsilon \right)^2} \Biggr] \frac{\total^d q}{(2 \pi)^d} \total \xi \frac{\total^d p}{(2 \pi)^d} + c \delta^4(x-y) \eqend{.}
\end{splitequation}
Analogously, we obtain
\begin{splitequation}
&\left[ \partial_\mu G^\mathrm{F}(x,y) \partial_\nu G^\mathrm{F}(x,y) \right]_\mathrm{ren} \\
&\quad= \frac{1}{4} \eta_{\mu\nu} \lim_{d \to 4} \int \mathe^{\mathi p (x-y)} \int_0^1 \int q^2 \Biggl[ \frac{1}{\left[ q^2 + \xi (1-\xi) p^2 + m^2 - \mathi \epsilon \right]^2} - \frac{1}{\left( q^2 + m^2 - \mathi \epsilon \right)^2} \\
&\hspace{14em}+ 2 \frac{\xi (1-\xi) p^2}{\left( q^2 + m^2 - \mathi \epsilon \right)^3} \Biggr] \frac{\total^d q}{(2 \pi)^d} \total \xi \frac{\total^d p}{(2 \pi)^d} \\
&\qquad- \lim_{d \to 4} \int \mathe^{\mathi p (x-y)} \int_0^1 \xi (1-\xi) p_\mu p_\nu \\
&\hspace{6em}\times \int \left[ \frac{1}{\left[ q^2 + \xi (1-\xi) p^2 + m^2 - \mathi \epsilon \right]^2} - \frac{1}{\left( q^2 + m^2 - \mathi \epsilon \right)^2} \right] \frac{\total^d q}{(2 \pi)^d} \total \xi \frac{\total^d p}{(2 \pi)^d} \\
&\qquad+ c_1 \partial_\mu \partial_\nu \delta^4(x-y) + c_2 \eta_{\mu\nu} \delta^4(x-y) + c_3 \eta_{\mu\nu} \partial^2 \delta^4(x-y) \eqend{,}
\end{splitequation}
where we also used rotational symmetry of the $q$ integral to simplify the result.

We evaluate the integrals explicitly, using that
\begin{equation}
\int \frac{(q^2)^a}{\left( q^2 + M - \mathi \epsilon \right)^b} \frac{\total^d q}{(2\pi)^d} = \frac{\mathi}{(4\pi)^\frac{d}{2}} \frac{\Gamma\left( a + \frac{d}{2} \right) \Gamma\left( b-a - \frac{d}{2} \right)}{\Gamma\left( \frac{d}{2} \right) \Gamma(b)} \left( M - \mathi \epsilon \right)^{\frac{d}{2}-b+a} \eqend{,}
\end{equation}
and taking the limit $d \to 4$ in the result. That is, we have
\begin{splitequation}
&\lim_{d \to 4} \int \left[ \frac{1}{\left( q^2 + M_1 - \mathi \epsilon \right)^2} - \frac{1}{\left( q^2 + M_2 - \mathi \epsilon \right)^2} \right] \frac{\total^d q}{(2\pi)^d} \\
&\quad= \frac{\mathi}{(4 \pi)^2} \Bigl[ \ln\left( M_2 - \mathi \epsilon \right) - \ln\left( M_1 - \mathi \epsilon \right) \Bigr]
\end{splitequation}
and
\begin{splitequation}
&\lim_{d \to 4} \int q^2 \left[ \frac{1}{\left( q^2 + M_1 - \mathi \epsilon \right)^2} - \frac{1}{\left( q^2 + M_2 - \mathi \epsilon \right)^2} - \frac{2 (M_2-M_1)}{\left( q^2 + M_2 - \mathi \epsilon \right)^3} \right] \frac{\total^d q}{(2\pi)^d} \\
&\quad= 2 \frac{\mathi}{(4 \pi)^2} \Bigl[ M_2 - M_1 + M_1 \ln\left( M_1 - \mathi \epsilon \right) - M_1 \ln\left( M_2 - \mathi \epsilon \right) \Bigr] \eqend{.}
\end{splitequation}
Afterwards, the $\xi$ integrals can also be computed, and we obtain
\begin{splitequation}
&\int_0^1 \ln\left[ 1 + \xi (1-\xi) \frac{p^2}{m^2} - \mathi \epsilon \right] \total \xi \\
&\quad= \left[ \frac{2\xi-1}{2} \ln\left[ 1 + \xi (1-\xi) \frac{4}{\sigma-1} \right] - 2 \xi + \frac{\sqrt{ \sigma }}{2} \ln\left( \frac{2 \xi - 1 + \sqrt{ \sigma }}{2 \xi - 1 - \sqrt{ \sigma }} \right) \right]_0^1 \\
&\quad= - 2 + \sqrt{ \sigma } \ln\left( \frac{1 + \sqrt{ \sigma }}{1 - \sqrt{ \sigma }} \right) = - 2 + 2 \sqrt{\sigma} \artanh\sqrt{\sigma} \eqend{,}
\end{splitequation}
where we defined
\begin{equation}
\sigma = 1 + 4 \frac{m^2}{p^2 - \mathi \epsilon} \eqend{,}
\end{equation}
and
\begin{splitequation}
&\int_0^1 \xi (1-\xi) \ln\left[ 1 + \xi (1-\xi) \frac{p^2}{m^2} - \mathi \epsilon \right] \total \xi \\
&\quad= \Biggl[ \frac{(2\xi-1) (1+2\xi-2\xi^2)}{12} \left[ \ln\left[ 1 + \xi (1-\xi) \frac{4}{\sigma-1} \right] - \frac{2}{3} \right] \\
&\hspace{4em}+ \frac{\sigma-2}{6} \xi + \frac{(3-\sigma) \sqrt{ \sigma }}{24} \ln\left( \frac{2 \xi - 1 + \sqrt{ \sigma }}{2 \xi - 1 - \sqrt{ \sigma }} \right) \Biggr]_0^1 \\
&\quad= \frac{\sigma}{6} - \frac{4}{9} + \frac{(3-\sigma) \sqrt{ \sigma }}{6} \artanh\sqrt{\sigma} \eqend{.}
\end{splitequation}

This then results in
\begin{equation}
\left[ G^\mathrm{F}(x,y) \right]^2_\mathrm{ren} = - \frac{\mathi}{8 \pi^2} I^\mathrm{F}_0(x,y) + c' \delta^4(x-y)
\end{equation}
and
\begin{splitequation}
\left[ \partial_\mu G^\mathrm{F}(x,y) \partial_\nu G^\mathrm{F}(x,y) \right]_\mathrm{ren} &= \frac{1}{6} \frac{\mathi}{(4 \pi)^2} \eta_{\mu\nu} m^2 \left[ I^\mathrm{F}_{-1}(x,y) + 4 I^\mathrm{F}_0(x,y) \right] \\
&\quad- \frac{1}{3} \frac{\mathi}{(4 \pi)^2} \partial_\mu \partial_\nu \left[ I^\mathrm{F}_0(x,y) - 2 I^\mathrm{F}_1(x,y) - 2 m^2 G^\mathrm{F}_0(x,y) \right] \\
&\quad+ c_1' \partial_\mu \partial_\nu \delta^4(x-y) + c_2' \eta_{\mu\nu} \delta^4(x-y) + c_3' \eta_{\mu\nu} \partial^2 \delta^4(x-y) \eqend{,}
\end{splitequation}
where we defined the basic integral
\begin{equation}
\label{eq:app_ifk_def}
I^\mathrm{F}_k(x,y) = \int \mathe^{\mathi p (x-y)} \left( \frac{m^2}{p^2 - \mathi \epsilon} \right)^k \sqrt{ 1 + 4 \frac{m^2}{p^2 - \mathi \epsilon} } \artanh\sqrt{ 1 + 4 \frac{m^2}{p^2 - \mathi \epsilon} } \frac{\total^4 p}{(2 \pi)^4} \eqend{,}
\end{equation}
set $G^\mathrm{F}_0(x,y) = G^\mathrm{F}(x,y) \rvert_{m^2 = 0}$, and absorbed the local term in the constants $c$ and $c_i$, defining
\begin{equations}
c' &= c + \frac{\mathi}{8 \pi^2} \eqend{,} \\
c_1' &= c_1 + \frac{5}{18} \frac{\mathi}{(4 \pi)^2} \eqend{,} \\
c_2' &= c_2 - \frac{2}{3} \frac{\mathi}{(4 \pi)^2} m^2 \eqend{,} \\
c_3' &= c_3 + \frac{2}{9} \frac{\mathi}{(4 \pi)^2} \eqend{.}
\end{equations}
Note that the basic integral $I^\mathrm{F}_k$~\eqref{eq:app_ifk_def} satisfies the recursion relation
\begin{equation}
\partial^2 I^\mathrm{F}_k(x,y) = - m^2 I^\mathrm{F}_{k-1}(x,y) \eqend{.}
\end{equation}

Taking everything together, it follows that
\begin{splitequation}
&2 \partial^\mu \left[ \partial_\mu G^\mathrm{F}(x,y) \partial_\nu G^\mathrm{F}(x,y) \right]_\mathrm{ren} - \partial_\nu \left[ \partial_\mu G^\mathrm{F}(x,y) \partial^\mu G^\mathrm{F}(x,y) \right]_\mathrm{ren} - m^2 \partial_\nu \left[ G^\mathrm{F}(x,y) \right]^2_\mathrm{ren} \\
&\quad= \partial_\nu \left[ ( c_1' - 2 c_3' ) \partial^2 \delta^4(x-y) - \left( 2 c_2' + c' m^2 - \frac{2}{3} \frac{\mathi}{(4 \pi)^2} m^2 \right) \delta^4(x-y) \right] \eqend{,}
\end{splitequation}
and to fulfill the condition~\eqref{eq:app_condition} we need to choose $c_1' = 2 c_3'$ and $c_2' = \frac{1}{3} \frac{\mathi}{(4 \pi)^2} m^2 - \frac{c'}{2} m^2$. This means that we define
\begin{equation}
\label{eq:app_gfren2}
\left[ G^\mathrm{F}(x,y) \right]^2_\mathrm{ren} = - \frac{\mathi}{8 \pi^2} I^\mathrm{F}_0(x,y) + c \delta^4(x-y)
\end{equation}
and
\begin{splitequation}
\label{eq:app_dgfren2}
\left[ \partial_\mu G^\mathrm{F}(x,y) \partial_\nu G^\mathrm{F}(x,y) \right]_\mathrm{ren} &= \frac{1}{6} \frac{\mathi}{(4 \pi)^2} \eta_{\mu\nu} m^2 \left[ I^\mathrm{F}_{-1}(x,y) + 4 I^\mathrm{F}_0(x,y) + 2 \delta^4(x-y) \right] \\
&\quad- \frac{1}{3} \frac{\mathi}{(4 \pi)^2} \partial_\mu \partial_\nu \left[ I^\mathrm{F}_0(x,y) - 2 I^\mathrm{F}_1(x,y) - 2 m^2 G^\mathrm{F}_0(x,y) \right] \\
&\quad- \frac{c}{2} m^2 \eta_{\mu\nu} \delta^4(x-y) + c' \left( 2 \partial_\mu \partial_\nu + \eta_{\mu\nu} \partial^2 \right) \delta^4(x-y) \eqend{,}
\end{splitequation}
where now $c$ and $c'$ are (new) constants that are ultimately fixed by mass and field strength renormalization. Similarly, using that
\begin{equation}
\label{eq:gp_def}
G^+(x,y) = - 2 \pi \mathi \int \mathe^{\mathi p (x-y)} \delta(p^2+m^2) \Theta(p^0) \frac{\total^d p}{(2 \pi)^d}
\end{equation}
we obtain
\begin{equation}
\left[ G^+(x,y) \right]^2 = - (2 \pi)^2 \int \mathe^{\mathi p (x-y)} \left[ \int \delta[ p^2 - 2 (pq) ] \Theta(p^0-q^0) \delta(q^2+m^2) \Theta(q^0) \frac{\total^4 q}{(2 \pi)^4} \right] \frac{\total^4 p}{(2 \pi)^4} \eqend{,}
\end{equation}
which is already a well-defined distribution in four dimensions, and we compute
\begin{splitequation}
&\int \delta[ p^2 - 2 (pq) ] \Theta(p^0-q^0) \delta(q^2+m^2) \Theta(q^0) \frac{\total^4 q}{(2 \pi)^4} \\
&= \frac{1}{4 \pi} \int \delta\left[ p^2 + 2 p^0 \sqrt{ \vec{q}^2 + m^2 } - 2 (\vec{p}\vec{q}) \right] \frac{\Theta(p^0 - \sqrt{ \vec{q}^2 + m^2 })}{\sqrt{ \vec{q}^2 + m^2 }} \frac{\total^3 \vec{q}}{(2 \pi)^3} \\
&= \frac{1}{16 \pi^3} \int_0^\infty \int_0^\pi \delta\left( p^2 + 2 p^0 \sqrt{ q^2 + m^2 } - 2 \abs{\vec{p}} q \cos \theta \right) \frac{\Theta(p^0 - \sqrt{ q^2 + m^2 })}{\sqrt{ q^2 + m^2 }} \sin \theta q^2 \total \theta \total q \\
&= \frac{1}{16 \pi^3} \int_0^\infty \int \frac{\Theta(1 - x^2)}{2 \abs{\vec{p}}} \delta\left( \frac{p^2 + 2 p^0 \sqrt{ q^2 + m^2 }}{2 \abs{\vec{p}} q} - x \right) \total x \frac{\Theta(p^0 - \sqrt{ q^2 + m^2 })}{\sqrt{ q^2 + m^2 }} q \total q \\
&= \frac{1}{16 \pi^3} \int_0^\infty \frac{\Theta\left( 4 p^2 ( q^2 + m^2 ) - ( p^2 )^2 - 4 \vec{p}^2 m^2 - 4 p^2 p^0 \sqrt{ q^2 + m^2 } \right)}{2 \abs{\vec{p}}} \frac{\Theta(p^0 - \sqrt{ q^2 + m^2 })}{\sqrt{ q^2 + m^2 }} q \total q \\
&= \frac{1}{16 \pi^3} \Theta(p^0-m) \int_m^{p^0} \frac{\Theta\left[ 4 p^2 k^2 - 4 p^2 m^2 - ( p^2 )^2 - 4 (p^0)^2 m^2 - 4 p^2 p^0 k \right]}{2 \abs{\vec{p}}} \total k \\
&= \frac{1}{64 \pi^3} \Theta(p^0-m) \int_{- \frac{p^0}{\abs{\vec{p}}}}^{\frac{p^0-2m}{\abs{\vec{p}}}} \Theta\left[ p^2 \ell^2 - ( p^2 + 4 m^2 ) \right] \total \ell \eqend{.}
\end{splitequation}
If $p^2 \geq 0$, so in particular $\abs{\vec{p}} \geq p^0$, we have $\ell^2 \leq 1$ and the Heaviside $\Theta$ function vanishes. If $p^2 \leq 0$ but $p^2 + 4 m^2 \geq 0$, the Heaviside $\Theta$ function vanishes as well. We thus obtain a $\Theta(-p^2-4m^2)$ and can divide the argument of the Heaviside $\Theta$ function by $-p^2$. It follows that $\ell^2 \leq 1 + \frac{4 m^2}{p^2}$, which implies in particular that $- \frac{p^0}{\abs{\vec{p}}} \leq \ell \leq \frac{p^0-2m}{\abs{\vec{p}}}$, and thus we obtain
\begin{equation}
\int \delta[ p^2 - 2 (pq) ] \Theta(p^0-q^0) \delta(q^2+m^2) \Theta(q^0) \frac{\total^4 q}{(2 \pi)^4} = \frac{1}{32 \pi^3} \Theta(p^0) \Theta(-p^2-4m^2) \sqrt{ 1 + \frac{4 m^2}{p^2} }
\end{equation}
and
\begin{equation}
\label{eq:app_gren2}
\left[ G^+(x,y) \right]^2 = - \frac{\mathi}{8 \pi^2} I^+_0(x,y) \eqend{.}
\end{equation}
where we defined the basic integral
\begin{equation}
\label{eq:app_ipk_def}
I^+_k(x,y) = - \mathi \pi \int \mathe^{\mathi p (x-y)} \left( \frac{m^2}{p^2} \right)^k \Theta(p^0) \Theta(-p^2-4m^2) \sqrt{ 1 + \frac{4 m^2}{p^2} } \frac{\total^4 p}{(2 \pi)^4} \eqend{.}
\end{equation}
This agrees (up to the sign of $p^0$) with the results in~\cite[App.~C]{martinverdaguer2000}, and the basic integral satisfies the recursion relation
\begin{equation}
\partial^2 I^+_k(x,y) = - m^2 I^+_{k-1}(x,y) \eqend{.}
\end{equation}

Analogously, it follows that
\begin{splitequation}
\label{eq:app_dgren2}
&\partial_\mu G^+(x,y) \partial_\nu G^+(x,y) \\
&\quad= (2 \pi)^2 \int \mathe^{\mathi p (x-y)} \left[ \int (p-q)_\mu q_\nu \delta[ p^2 - 2 (pq) ] \Theta(p^0-q^0) \delta(q^2+m^2) \Theta(q^0) \frac{\total^4 q}{(2 \pi)^4} \right] \frac{\total^4 p}{(2 \pi)^4} \\
&\quad= \frac{\pi^2}{3} \int \mathe^{\mathi p (x-y)} \left[ 2 \frac{p_\mu p_\nu}{p^2} ( p^2 - 2 m^2 ) + \eta_{\mu\nu} ( p^2 + 4 m^2 ) \right] \\
&\qquad\qquad\times \int \delta[ p^2 - 2 (pq) ] \Theta(p^0-q^0) \delta(q^2+m^2) \Theta(q^0) \frac{\total^4 q}{(2 \pi)^4} \frac{\total^4 p}{(2 \pi)^4} \\
&\quad= \frac{1}{96 \pi} \int \mathe^{\mathi p (x-y)} \left[ 2 \frac{p_\mu p_\nu}{p^2} ( p^2 - 2 m^2 ) + \eta_{\mu\nu} ( p^2 + 4 m^2 ) \right] \\
&\qquad\qquad\times \Theta(p^0) \Theta(-p^2-4m^2) \sqrt{ 1 + \frac{4 m^2}{p^2} } \frac{\total^4 p}{(2 \pi)^4} \\
&\quad= \frac{1}{6} \frac{\mathi}{(4 \pi)^2} \eta_{\mu\nu} m^2 \left[ I^+_{-1}(x,y) + 4 I^+_0(x,y) \right] - \frac{1}{3} \frac{\mathi}{(4 \pi)^2} \partial_\mu \partial_\nu \left[ I^+_0(x,y) - 2 I^+_1(x,y) \right] \eqend{.}
\end{splitequation}
The analogous equations hold with $G^-$, where $I^-_k$ contains $\Theta(-p^0)$ instead of $\Theta(p^0)$.

For the combination relevant to the stress tensor, we then obtain
\begin{splitequation}
\label{eq:app_gf_stress}
&\left[ \partial_\mu G^\mathrm{F}(x,y) \partial_\nu G^\mathrm{F}(x,y) \right]_\mathrm{ren} - \frac{1}{2} \eta_{\mu\nu} \left[ \partial_\rho G^\mathrm{F}(x,y) \partial^\rho G^\mathrm{F}(x,y) \right]_\mathrm{ren} - \frac{1}{2} \eta_{\mu\nu} m^2 \left[ G^\mathrm{F}(x,y) \right]^2_\mathrm{ren} \\
&= - \frac{1}{3} \frac{\mathi}{(4 \pi)^2} \left( \partial_\mu \partial_\nu - \eta_{\mu\nu} \partial^2 \right) \left[ I^\mathrm{F}_0(x,y) - 2 I^\mathrm{F}_1(x,y) - 2 m^2 G^\mathrm{F}_0(x,y) \right] \\
&\quad+ 2 c' \left( \partial_\mu \partial_\nu - \eta_{\mu\nu} \partial^2 \right) \delta^4(x-y)
\end{splitequation}
and
\begin{splitequation}
\label{eq:app_gpm_stress}
&\partial_\mu G^\pm(x,y) \partial_\nu G^\pm(x,y) - \frac{1}{2} \eta_{\mu\nu} \partial_\rho G^\pm(x,y) \partial^\rho G^\pm(x,y) - \frac{1}{2} \eta_{\mu\nu} m^2 \left[ G^\pm(x,y) \right]^2 \\
&= - \frac{1}{3} \frac{\mathi}{(4 \pi)^2} \left( \partial_\mu \partial_\nu - \eta_{\mu\nu} \partial^2 \right) \left[ I^\pm_0(x,y) - 2 I^\pm_1(x,y) \right] \eqend{.}
\end{splitequation}

\section{Covariance of the commutator and retarded Green's function}
\label{app:cov}

Here, we show that the commutator function and the retarded and advanced propagators are covariant under the action of the differential operator~\eqref{eq:l_def}
\begin{equation}
L f(x) = x^0 \partial_1 f(x) + x^1 \partial_0 f(x) \eqend{.}
\end{equation}
For this, we employ the explicit expression~\eqref{eq:gretadv_def} and compute
\begin{splitequation}
\left( L_x + L_y \right) \Delta(x,y) &= \left( x^0 \frac{\partial}{\partial x^1} + x^1 \frac{\partial}{\partial x^0} \right) \Delta(x,y) + \left( y^0 \frac{\partial}{\partial y^1} + y^1 \frac{\partial}{\partial y^0} \right) \Delta(x,y) \\
&= - 2 \pi \mathi \int \delta(p^2+m^2) \sgn(p^0) \left[ (x^0-y^0) \mathi p^1 - (x^1-y^1) \mathi p^0 \right] \mathe^{\mathi p (x-y)} \frac{\total^d p}{(2\pi)^d} \\
&= - 2 \pi \mathi \int \delta(p^2+m^2) \sgn(p^0) \left( - p^1 \frac{\partial}{\partial p^0} - p^0 \frac{\partial}{\partial p^1} \right) \mathe^{\mathi p (x-y)} \frac{\total^d p}{(2\pi)^d} \\
&= - 2 \pi \mathi \int \mathe^{\mathi p (x-y)} \left( p^1 \frac{\partial}{\partial p^0} + p^0 \frac{\partial}{\partial p^1} \right) \left[ \delta(p^2+m^2) \sgn(p^0) \right] \frac{\total^d p}{(2\pi)^d} \\
&= - 2 \pi \mathi \int \mathe^{\mathi p (x-y)} 2 p^1 \delta(\omega_\vec{p}^2) \delta(p^0) \frac{\total^d p}{(2\pi)^d} = 0 \eqend{,}
\end{splitequation}
since $\omega_\vec{p}^2 \geq m^2 > 0$ in the massive case, and $p^1 \delta(\vec{p}^2) = 0$ in the massless case.

Using that $G^\mathrm{ret}(x,y) = \Theta(x^0-y^0) \Delta(x,y)$, we also compute
\begin{splitequation}
\left( L_x + L_y \right) G^\mathrm{ret}(x,y) &= \left( x^1 \frac{\partial}{\partial x^0} + y^1 \frac{\partial}{\partial y^0} \right) \Theta(x^0-y^0) \Delta(x,y) \\
&= (x^1-y^1) \delta(x^0-y^0) \Delta(x,y) = 0 \eqend{,}
\end{splitequation}
where the last equality follows from~\eqref{eq:delta_identities}, and analogously $\left( L_x + L_y \right) G^\mathrm{adv}(x,y) = 0$. It follows that for any compactly supported function $f$ we have
\begin{splitequation}
\label{eq:app_l_commute_delta}
( \Delta L f )(x) &= \int \Delta(x,y) L f(y) \total^4 y = - \int L_y \Delta(x,y) f(y) \total^4 y \\
&= \int L_x \Delta(x,y) f(y) \total^4 y = L ( \Delta f )(x) \eqend{,}
\end{splitequation}
and analogously
\begin{equation}
\label{eq:app_l_commute_gretadv}
( G^\mathrm{ret/adv} L f )(x) = L ( G^\mathrm{ret/adv} f )(x) \eqend{.}
\end{equation}

\section{Extended Green's identity}
\label{app:green}

Consider two smooth functions $f$ and $g$, where $g$ has compact support but $f$ not. In this constellation, we first split
\begin{equation}
\int f(x) ( \Delta g )(x) \total^4 x = \int_{x^0 \leq t} f(x) ( \Delta g )(x) \total^4 x + \int_{x^0 \geq t} f(x) ( \Delta g )(x) \total^4 x \eqend{,}
\end{equation}
and then use that $G^\mathrm{ret/adv}(x,y)$ are fundamental solutions of the Klein--Gordon equation~\eqref{eq:gretadv_fundamentalsol} to obtain
\begin{splitequation}
&\int_{x^0 \leq t} f(x) ( \Delta g )(x) \total^4 x \\
&\quad= \int_{x^0 \leq t} \left( \partial^2 - m^2 \right) ( G^\mathrm{ret} f )(x) ( \Delta g )(x) \total^4 x \\
&\quad= - \int_{x^0 \leq t} \partial_0 \left[ \partial_0 ( G^\mathrm{ret} f )(x) ( \Delta g )(x) \right] \total^4 x + \int_{x^0 \leq t} \partial_0 ( G^\mathrm{ret} f )(x) \partial_0 ( \Delta g )(x) \total^4 x \\
&\qquad+ \int_{x^0 \leq t} \left( \partial^k \partial_k - m^2 \right) ( G^\mathrm{ret} f )(x) ( \Delta g )(x) \total^4 x \\
&\quad= \left( \lim_{T \to - \infty} \int_{x^0 = T} - \int_{x^0 = t} \right) \partial_0 ( G^\mathrm{ret} f )(x) ( \Delta g )(x) \total^3 \vec{x} \\
&\qquad+ \int_{x^0 \leq t} \partial_0 ( G^\mathrm{ret} f )(x) \partial_0 ( \Delta g )(x) \total^4 x + \int_{x^0 \leq t} ( G^\mathrm{ret} f )(x) \left( \partial^k \partial_k - m^2 \right) ( \Delta g )(x) \total^4 x \\
&\quad= \left( \lim_{T \to - \infty} \int_{x^0 = T} - \int_{x^0 = t} \right) \partial_0 ( G^\mathrm{ret} f )(x) ( \Delta g )(x) \total^3 \vec{x} \\
&\qquad+ \int_{x^0 \leq t} \partial_0 \left[ ( G^\mathrm{ret} f )(x) \partial_0 ( \Delta g )(x) \right] \total^4 x \\
&\quad= \left( \lim_{T \to - \infty} \int_{x^0 = T} - \int_{x^0 = t} \right) \left[ \partial_0 ( G^\mathrm{ret} f )(x) ( \Delta g )(x) - ( G^\mathrm{ret} f )(x) \partial_0 ( \Delta g )(x) \right] \total^3 \vec{x} \eqend{,}
\end{splitequation}
where in the third equality we used that $\Delta g$ is spatially compact to integrate spatial derivatives by parts, and the fourth that $\Delta$ is a bisolution of the Klein--Gordon equation. Analogously, one obtains
\begin{splitequation}
&\int_{x^0 \geq t} f(x) ( \Delta g )(x) \total^4 x \\
&\quad= - \left( \lim_{T \to \infty} \int_{x^0 = T} - \int_{x^0 = t} \right) \left[ \partial_0 ( G^\mathrm{adv} f )(x) ( \Delta g )(x) - ( G^\mathrm{adv} f )(x) \partial_0 ( \Delta g )(x) \right] \total^3 \vec{x} \eqend{,}
\end{splitequation}
and using that $\Delta = G^\mathrm{ret} - G^\mathrm{adv}$ the extended Green's identity
\begin{splitequation}
\label{eq:app_green_spacecompact}
\int f(x) ( \Delta g )(x) \total^4 x &= \int_{x^0 = t} \left[ ( \Delta f )(x) \partial_0 ( \Delta g )(x) - \partial_0 ( \Delta f )(x) ( \Delta g )(x) \right] \total^3 \vec{x} \\
&\quad+ \lim_{T \to \infty} \int_{x^0 = T} \left[ ( G^\mathrm{adv} f )(x) \partial_0 ( \Delta g )(x) - \partial_0 ( G^\mathrm{adv} f )(x) ( \Delta g )(x) \right] \total^3 \vec{x} \\
&\quad- \lim_{T \to - \infty} \int_{x^0 = T} \left[ ( G^\mathrm{ret} f )(x) \partial_0 ( \Delta g )(x) - \partial_0 ( G^\mathrm{ret} f )(x) ( \Delta g )(x) \right] \total^3 \vec{x}
\end{splitequation}
for some arbitrary $t \in \mathbb{R}$.

If $f$ has compact support, $( G^\mathrm{ret} f )(x)$ vanishes for all $x \not\in J^+( \supp f )$, and in particular in the limit $t \to - \infty$. Analogously, $( G^\mathrm{adv} f )(x)$ vanishes for all $x \not\in J^-( \supp f )$, and in particular in the limit $t \to \infty$. Therefore, the last two integrals in~\eqref{eq:app_green_spacecompact} vanish, and we obtain the classical Green's identity
\begin{equation}
\label{eq:app_green_identity}
\int f(x) ( \Delta g )(x) \total^4 x = \int_{x^0 = t} \Bigl[ ( \Delta f )(x) \partial_0 ( \Delta g )(x) - ( \Delta g )(x) \partial_0 ( \Delta f )(x) \Bigr] \total^3 \vec{x} \eqend{.}
\end{equation}

\bibliography{literature}

\end{document}